\documentclass[aps,pra,reprint,twocolumn,showpacs,longbibliography,superscriptaddress]{revtex4-1}
\usepackage{amsmath,amssymb,graphicx,units,hyperref,hypernat}
\usepackage[usenames,dvipsnames]{color}

\begin{document}
\title{Magnetism, transport and thermodynamics in two-dimensional half-filled Hubbard superlattices}
\author{Rubem Mondaini}
\affiliation{Beijing Computational Science Research Center, Beijing 100193, China}
\author{Thereza Paiva}
\affiliation{Instituto de F\'\i sica, Universidade Federal do Rio de Janeiro Cx.P. 68.528, 21941-972 Rio de Janeiro RJ, Brazil}

\begin{abstract}
We study magnetic, transport and thermodynamic properties of the half-filled two-dimensional ($2D$) Hubbard model with layered distributed repulsive interactions using unbiased finite temperature quantum Monte Carlo simulations. Antiferromagnetic long-ranged correlations at $T=0$ are confirmed by means of the magnetic structure factor and the onset of short-ranged ones is at a minimum temperature, which can be obtained by peaks in susceptibility and specific heat following a random-phase-approximation (RPA) prediction. We also show that transport is affected in the large interaction limit and is enhanced in the non-repulsive layers suggesting a change of dimensionality induced by increased interactions. Lastly, we show that by adiabatically switching the interactions in layered distributed patterns reduces the overall temperature of the system with a potential application in cooling protocols in cold atoms systems.
\end{abstract}
\maketitle

\section{Introduction}

Recent improvements on deposition techniques has enabled the growth of atomically precise layer sequences of different materials~\cite{Chakhalian2014}. Among the recently synthesized structures, the transition metal oxide superlattices (SL's), for example, play a key role, as they offer the potential for future use in devices~\cite{review-science}. LaAlO$_3$-SrTiO$_3$ superlattices, for instance, have been used to fabricate diodes with with room-temperature breakdown voltages of up to 200 V~\cite{Jany2010}, as well as field-effect devices~\cite{Forg2012, Zhou2013, Chen2016}. Most of the compounds used in these superlattices are characterized by the presence of strong electronic correlations, that give rise to complex collective quantum phases. Among the correlation-driven phenomena occurring in these materials one can highlight the interface superconductivity~\cite{Reyren2007}, magnetism between non-magnetic interfaces~\cite{Brinkman2007}, coexistence of magnetic order and two-dimensional superconductivity at LaAlO$_3$/SrTiO$_3$ interfaces~\cite{Li2011}, and others. These phenomena have led to the intense study of the {\it interface} between oxides in superlattice structures~\cite{review-oxides, review-science}.

Another interesting point of view is the study of the change of magnetic and transport properties as the {\it width} of one or both of the layers on a superlattice is altered. 
Superlattices made of paramagnetic correlated metal LaNiO$_3$ and wide-gap insulator LaAlO$_3$, grown by pulsed laser deposition, show collective metal-insulator transitions and antiferromagnetic transitions as a function of temperature when the lanthanum nickelate is as thin as two unit cells. Conversely, samples with thicker LaAlO$_3$ layers remain metallic and paramagnetic at all temperatures~\cite{Boris2011}. It is also possible to tune the magnetic character from antiferrogmagnetic to ferromagnetic of a thin film of LaAlO$_3$ grown on top of SrTiO$_3$ when the thickness of the the lanthanum alluminate has six or more unit-cells~\cite{Wang2015}.
Superlattices with heavy fermion compounds also show interesting behavior with decreasing layer thickness: epitaxially grown superlattices of antiferromagnetic  CeIn$_3$ and metallic LaIn$_3$~\cite{Shishido2010} show a linear  decrease  of the N\'eel temperature when the width of the CeIn$_3$ layer is reduced, vanishing when it is two atoms thick.

In the context of cold atoms, which presents a framework to investigate many-body phenomena in a highly tunable fashion~\cite{Bloch08, Bloch2012}, although spatially varying interactions have not yet been realized in optical lattices, they have recently become available in trapped ultracold gases. The ability to control a magnetic Feshbach resonance with laser light~\cite{Bauer2009}, has increased the tunability of interactions for bosonic systems. Submicrometer spatial modulation of the interaction was already achieved in a $^{174}$Yb Bose-Einstein condensate~\cite{boson_sl} and new optical controls of Feshbach resonances for fermionic ultracold gases~\cite{fermion_sl_1,fermion_sl_2,Jagannathan2016} have also been proposed. Only recently~\cite{Clark2015}, however, it was possible to overcome two major difficulties that plagues the experiments using optical Feshbach resonances: heating from off-resonant light scattering that leads to a rapid decay of the quantum gas and an unwanted shift of the energy levels that leads to the deformation of the trap potential, recently demonstrated by trapping a Bose-Einstein condensate of Cs atoms subjected to a position dependent modulation of the inter-atomic interactions.

Overall, either in condensed matter or in atomic physics realms, these experiments illustrate that the properties of otherwise homogeneous systems can be drastically altered when they are cast into ultra-thin layers forming a superlattice. A simple model that incorporates both fermionic correlations and superlattices structure can help to elucidate some of the issues in these fields and potentially indicate new routes of experimental investigation.  Here we study a two-dimensional model where one-dimensional strongly correlated  strips of width $L_{\rm U}$ are intercalated by non-interacting strips of width $L_0$, forming 2D superlattices.  The two-dimensional ``bulk'' non-interacting system corresponds to a paramagnetic metal; conversely, at half-filling, the two-dimensional interacting system has an antiferromagnetic, Mott-insulator ground-state.  The questions that we wish to address here are the following. (i)  What are the magnetic and  transport properties of these superlattices? (ii) How do they depend on layer thicknesses? (iii)  Is the magnetic order preserved in the presence of non-interacting sites?  (iv) How are the temperature scales affected by the superlattice structure? (v) Can we devise new cooling protocols in optical lattices by adiabatically changing the spatially modulated interactions?

The paper is organized as follows: Section~\ref{sec:M&M} describes the model and method used to perform the simulations. Section~\ref{sec:magnetism} describes magnetism whereas Sec.~\ref{sec:transport} investigates the resulting transport in these superlattice structures. Section~\ref{sec:tdynprop} is dedicated to the thermodynamical properties where the signatures of charge and spin fluctuations are analyzed in specific heat, spin susceptibility and entropy data; Sec.~\ref{sec:concl} summarizes our findings.

\section{Model and Method}
\label{sec:M&M}
\begin{figure}[!tb] 
  \includegraphics[width=0.75\columnwidth]{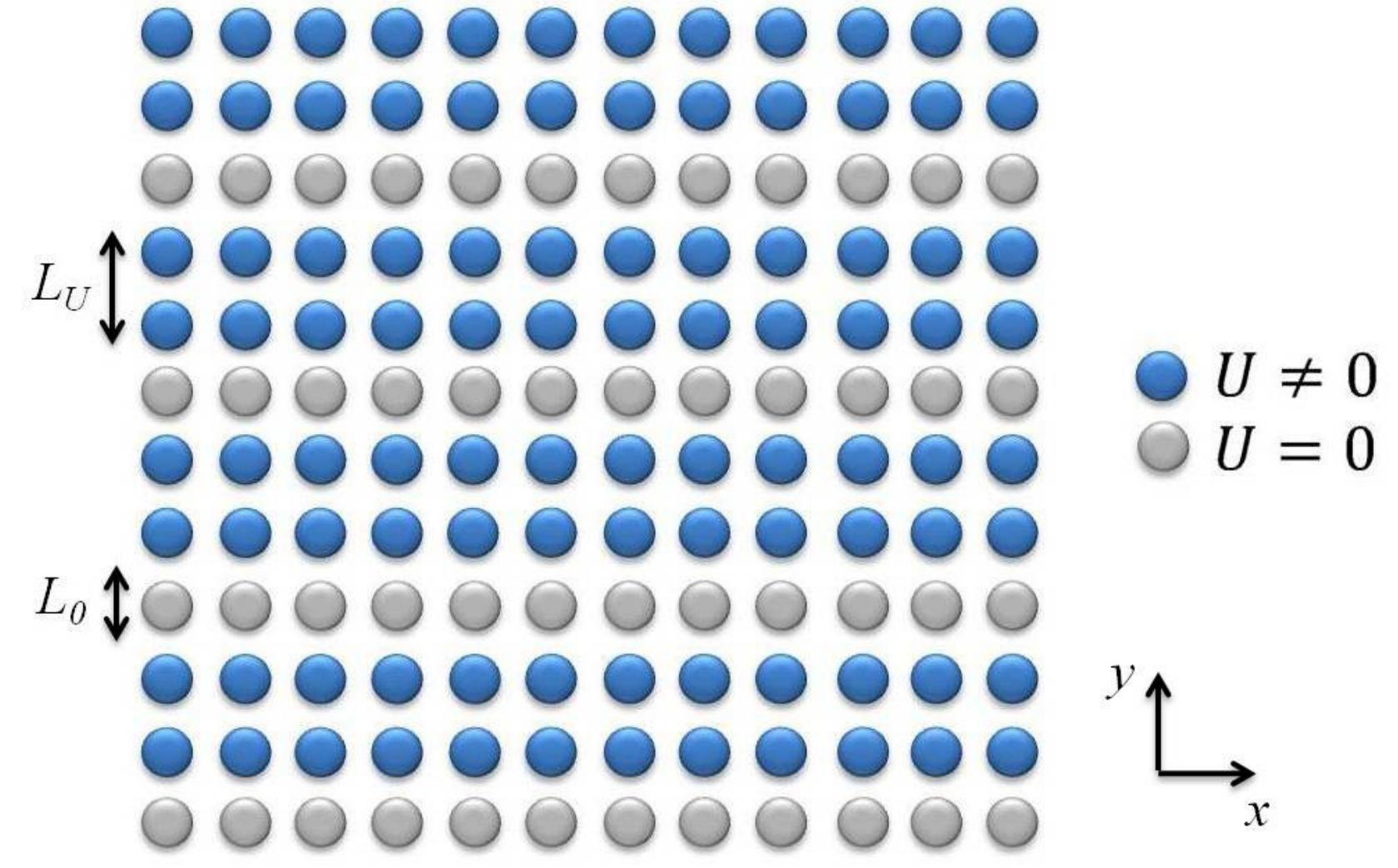}
 \vspace{-0.1cm}
 \caption{(Color online) Cartoon of the regularly distributed onsite repulsive interactions for the $L_U=2$ $L_0=1$ case in a $12\times12$ lattice. Throughout this work, $x$ represents the direction along the layers and $y$ perpendicular to it.}
 \label{fig:sl}
\end{figure}

We consider a modified version of the Hubbard Model (HM) with site-dependent repulsive interactions; the Hamiltonian, using periodic boundary conditions, reads:
\begin{eqnarray}
\mathcal{\hat H} &=& - t \sum_{\langle i,j\rangle,\sigma}
(\hat c^\dagger_{i\sigma} \hat c_{j \sigma}^{\phantom\dagger} +
\hat c^\dagger_{j\sigma} \hat c_{i \sigma}^{\phantom\dagger} )
\nonumber
\\
&&+ \sum_i U_i\left(\hat n_{i\uparrow}-\frac{1}{2}\right)\left(\hat n_{i\downarrow}-\frac{1}{2}\right)
-\mu \sum_{i,\sigma} \hat n_{i\sigma},
\label{eq:hamil}
\end{eqnarray}
where $\hat c^\dagger_{i\sigma}$($c_{i \sigma}^{\phantom\dagger}$) is the fermionic creation(annihilation) operator in site $i$ with pseudospin $\sigma=\uparrow,\downarrow$ and $\hat n_{i\sigma}$ is the number operator. $t$ is the hopping parameter between nearest neighbor sites $(\langle i,j\rangle)$, of an $L\times L$ square lattice, $U_i$ is the site dependent repulsion, and $\mu$ is the chemical potential that controls the band filling yielding a given electronic density $\rho$. The interaction term is written in particle-hole symmetric form. Thus, tuning $\mu=0$ drives the electronic occupation to one in all sites for any combination of the Hamiltonian parameters $t$, $U_i$ and temperatures $T$~\footnote{It is worth mentioning that the reason behind it is that in Eq.~(\ref{eq:hamil}) the repulsive sites ($U_i>0$) have a lower on-site energy which then balances the occupancy irrespectively of whether a site possesses $U=0$ or $U\neq0$. Hence, we are focused here on the effects of the underlying superlattice structure in the physical properties for a situation where charges are still homogeneously distributed.}. We have restricted our study to half-filled systems ($\rho=1.0$). To simulate the layered systems we construct a pattern of repulsive and non-interacting layers where $U>0$  and $U=0$, respectively. We define the width of the repulsive layer as $L_{\rm U}$ and the width of the non-interacting one as $L_0$ as depicted in Fig.~\ref{fig:sl} for the  $L_{\rm U}=2$ $L_0=1$ case, in a $12\times12$ lattice. Note that not all patterns are commensurate with the available lattice sizes we can numerically investigate (we have considered lattices up to $18 \times 18$), therefore, a finite size scaling analysis is in some cases elusive. We set $t$ as our energy scale.

The ground-state magnetic and transport properties of the site-dependent Hubbard model have been extensively studied in one-dimension. The non-symmetric Hamiltonian was studied with different numerical approaches, such as  the Lanczos method~\cite{TCLP96, TCLP98, TCLP00}, density matrix renormalization group~\cite{Malvezzi02}, within the Hartree-Fock approximation~\cite{Chowdhury07}, and within the Luttinger liquid framework~\cite{Silva_Valencia01}. The effect of an on-site energy in one of the sublattices was also considered~\cite{Gora98, Silva05} and was shown to strongly alter the ground state properties.

Going beyond one-dimensional systems, other studies have focused on the interface properties of metallic and interacting regions at finite temperatures. These studies often focus on the penetration of the magnetism in the metallic regions and induced metallic behavior on the insulating side in two-~\cite{Jiang12} and three dimensions~\cite{Euverte12,Zujev13}, where hybridization effects are explored by tuning the hopping at the interfaces in order to explore the interplay of magnetism and Kondo screening. Here, on the other hand, we aim to provide an in-depth study for the case of many interfaces forming a superlattice. Other studies~\cite{Zhong15} were primarily attained to the interface effects of metallic and insulating thin films with a potential realization of correlated transistors. We focus as well in the induced magnetism in metallic regions, the anisotropic transport due to the \textit{layered} structure, finite temperature scales for spin and moment formation and, lastly, in cooling mechanisms that could be potentially employed in cold atoms experiments in optical lattices.

We use finite temperature determinantal quantum Monte Carlo (QMC) simulations to unbiasedly probe magnetic, transport and thermodynamic properties of the half-filled two-dimensional superlattices. In this method, the partition function is expressed as a path integral by using the Suzuki-Trotter decomposition of $\exp(-\beta\mathcal{H})$, introducing the imaginary-time interval $\Delta\tau$.
The interaction term is decoupled through a discrete Hubbard-Stratonovich transformation~\cite{Blankenbecler81,Hirsch83}, which introduces an auxiliary Ising field.
This allows one to eliminate the fermionic degrees of freedom, and the summation over the auxiliary field (which depends both on the site and the imaginary time) is carried out stochastically.
Initially this field is generated randomly, and a local flip is attempted, with the acceptance rate given by the Metropolis algorithm.
The process of traversing the entire space-time lattice trying to change the auxiliary field variable constitutes one QMC sweep.
The errors associated with the Suzuki-Trotter decomposition in the QMC method are proportional to $\mathcal{O}\left[(\Delta\tau)^2 U\right]$, so we have set $\Delta\tau=0.125/t$ to $\Delta\tau=0.05/t$
depending on the value of the interaction strength $U$, to guarantee small enough systematic errors.~\footnote{We have typically used 1000 warm up sweeps for equilibration, followed by 4000 measuring sweeps, where the error bars are estimated by the statistical fluctuations. When necessary, the data were estimated over an average of simulations with different random seeds.}

\section{Magnetism}
\label{sec:magnetism}
\subsection{Short-ranged correlations}

The longstanding question of induced magnetism in metallic non-ordered regions due to the proximity to a magnetically ordered insulator can be initially tackled in a superlattice construction by investigating how the short-ranged correlations are modified by the layered pattern. Pushing the limits of short ranged to local (i.e., in the same site) we first investigate the local moment, defined as $\langle \left(\hat m_\mathbf{i}^z\right)^2\rangle \equiv \langle \left( \hat n_{\mathbf{i}\uparrow}-\hat n_{\mathbf{i}\downarrow}\right)^2\rangle$. Beyond its purely theoretical relevance, we stress that in the situation that the proposed Hamiltonian [Eq.~(\ref{eq:hamil})] could be emulated in optical lattices experiments, this is precisely the quantity that was recently measured to probe local spin order in a study of the two-dimensional Fermi-Hubbard model using trapped cold atoms~\cite{Cheuk1260}.   
From the theoretical point of view it is important to understand the local magnetic properties when approaching the ground state at $T=0$. However, with the connection to experiments in cold atoms in mind, again,  
here we will focus in ranges of temperatures that, although lower than the ones achieved in current experiments, could be potentially used as a guidance for future experiments.

\begin{figure}[!tb] 
  \includegraphics[width=0.99\columnwidth]{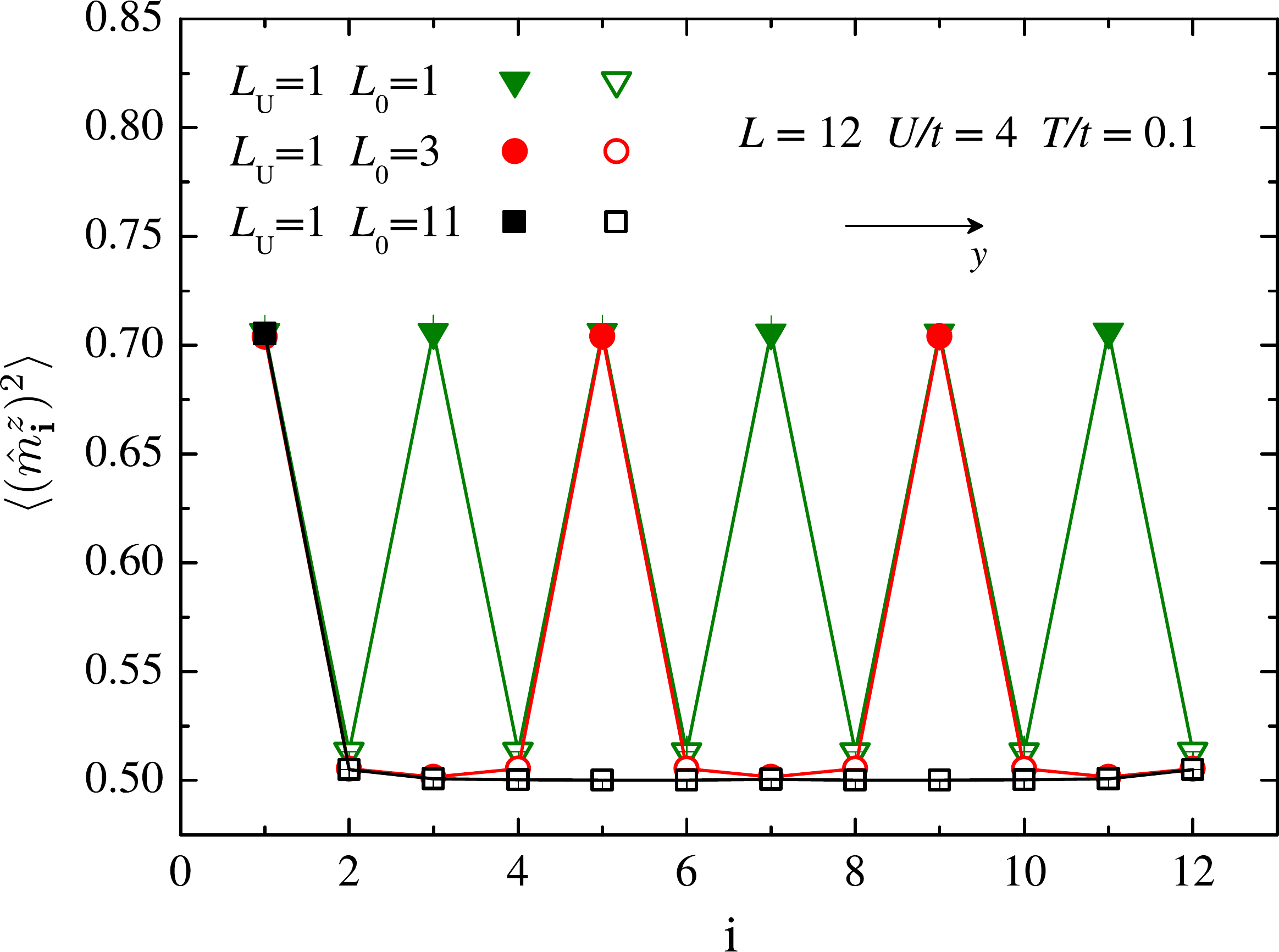}
 \caption{(Color online) Local moment profile throughout layers for different SL's with $L=12$ and $U/t=4$ at temperature $T/t=0.1$. Filled and empty symbols denote repulsive and free $(U=0)$ sites, respectively.}
 \label{fig:si2throughlayers}
\end{figure}

As we have discussed in the previous section, the symmetric form of the Hamiltonian requires that at half-filling the charge distribution is homogeneous throughout the lattice.  The local moment profile, on the other hand, is not homogeneous and strongly depends on the superlattice pattern, as can be clearly observed in Fig.~\ref{fig:si2throughlayers}, following closely the same periodicity of the superlatice structure\footnote{We focus on relatively small lattices for the current computational capabilities since the size-dependence is minimal for local quantities, as the local moment. This allows us to have very good statistics after all.}. Double occupations are less likely on repulsive sites than on non-interacting ones, therefore the local moment is larger on the interacting sites. In the homogeneous Hubbard model the local moment  increases monotonically  with the interaction strength~\cite{Hirsch85}. At half-filling for $U=0$, it takes the uncorrelated value $\langle \left(\hat m_\mathbf{i}^z\right)^2\rangle=1/2$, while as $U$ increases the double and empty occupancies decrease, until for $U\rightarrow\infty$ they are completely suppressed leading to $\langle \left(\hat m_\mathbf{i}^z\right)^2\rangle=1$, which corresponds to the spin-$\frac{1}{2}$ Heisenberg limit.

\begin{figure}[!tb] 
  \includegraphics[width=0.99\columnwidth]{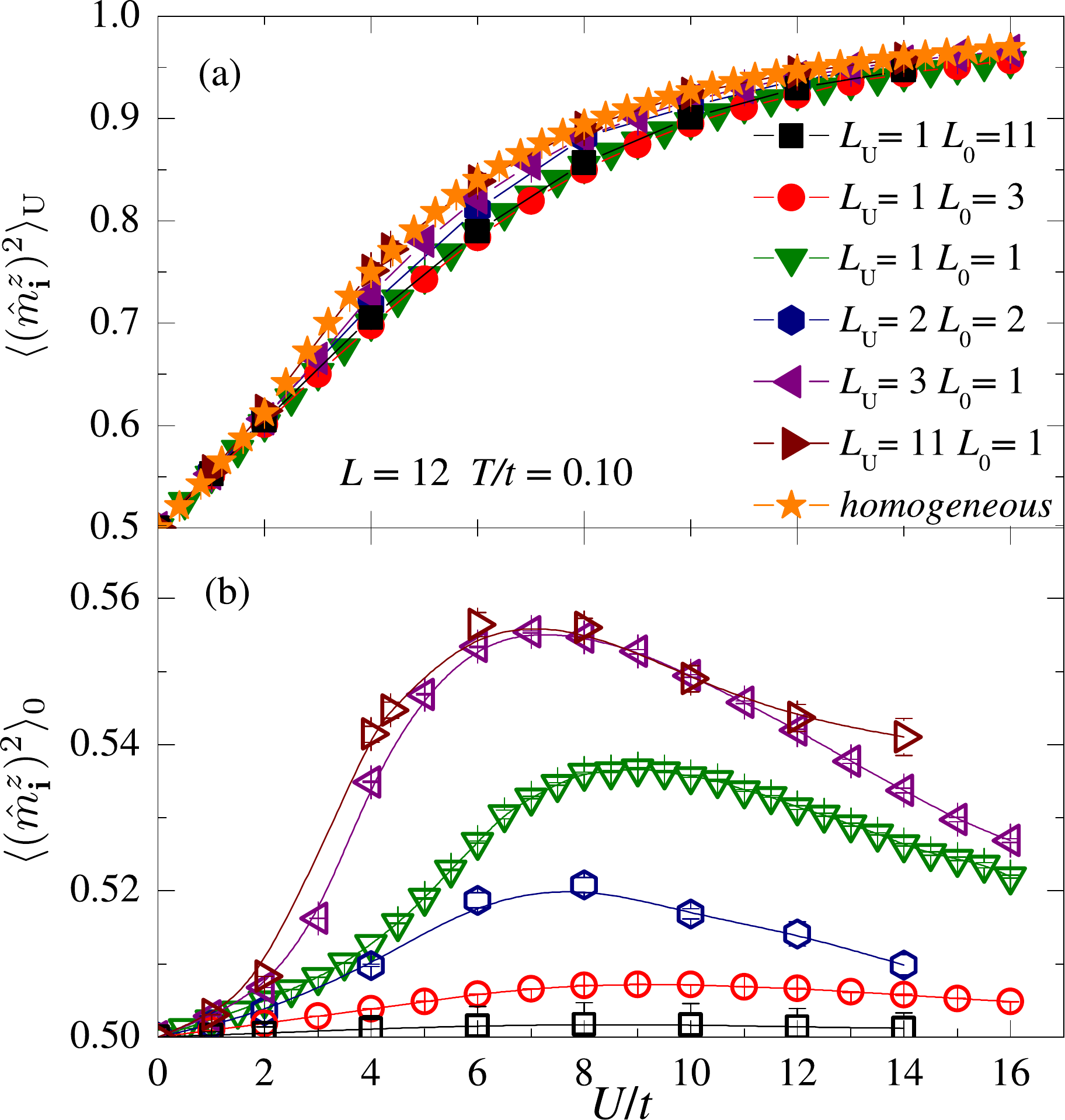}
 \caption{(Color online) Local moment dependence with $U/t$ for different SL's with $L=12$ at temperature $T/t=0.1$. Filled and empty symbols denote repulsive and free $(U=0)$ sites, respectively.}
 \label{fig:mi2vsU}
\end{figure}

\begin{figure}[!tb] 
  \includegraphics[width=1.0\columnwidth]{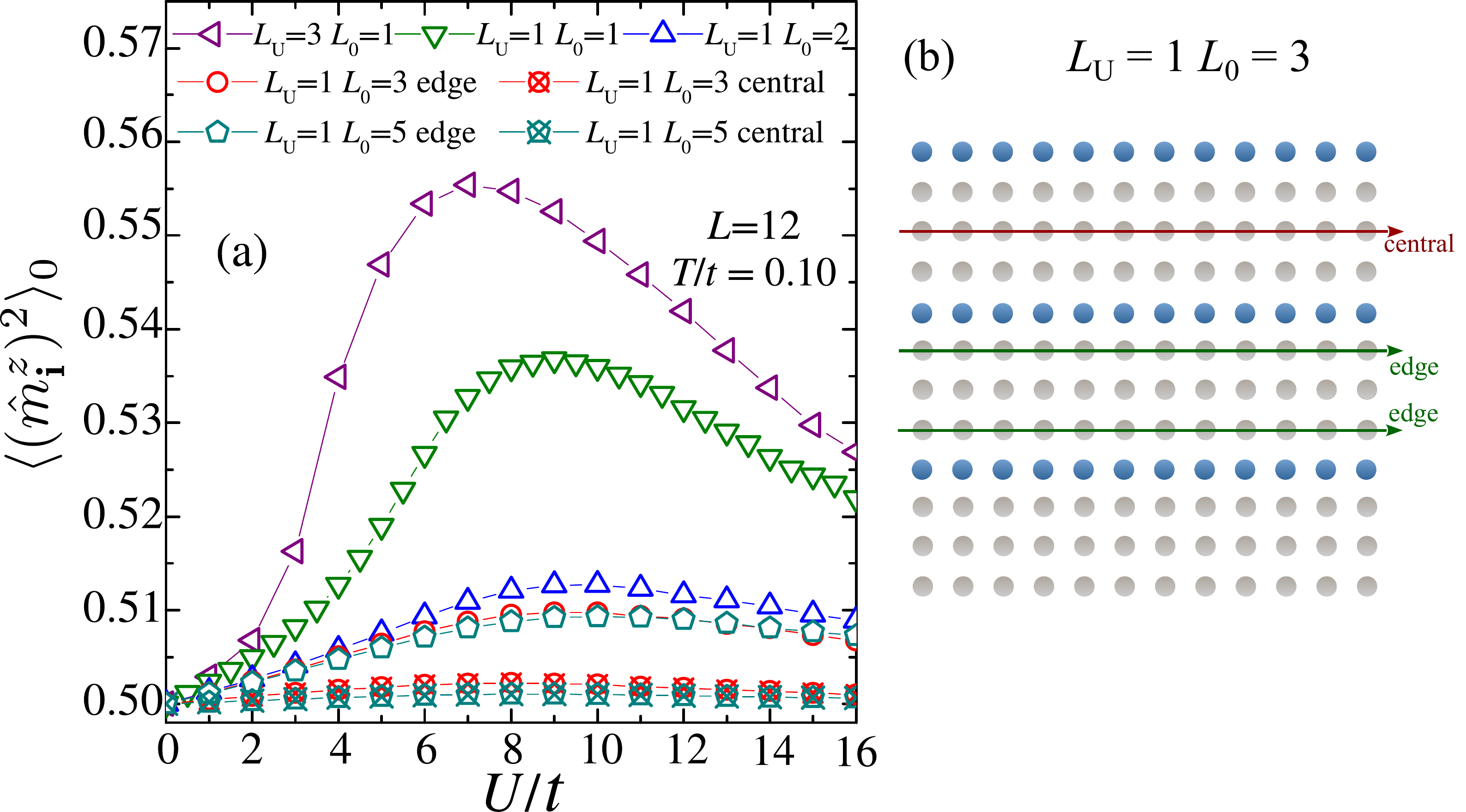}
 \caption{(Color online) (a) Local moment in free sites as a function of $U/t$ for different SL's with $L=12$ at temperature $T/t=0.1$. Empty symbols represent sites at the edge of the free layers whereas crossed symbols represent sites at the central line of the free layer: see cartoon in (b) with the example for the SL configuration $L_{\rm U} = 1 \,\, L_0=3$.}
 \label{fig:si20}
\end{figure}
 
The $U$-dependence of the local moment for the  layered system at $T/t=0.10$  is shown in Fig.~\ref{fig:mi2vsU}. The plot displays the  average local moment within repulsive [Fig.~\ref{fig:mi2vsU}(a)] and free  [Fig.~\ref{fig:mi2vsU}(b)] sites. In the former, $\langle \left(\hat m_\mathbf{i}^z\right)^2\rangle_{\rm U}$ increases monotonically with $U$, and approaches the values for the homogeneous system as $L_{\rm U}$  increases. On the other hand, the local moment in the non-interacting sites is affected by the strength of the interaction on the neighboring repulsive ones and displays a non-monotonic behavior with $U$. Starting from $U=0$, when we increase $U$, the effect of the interactions ``leaks" into the non-interacting sites, resulting in an increase of the local moment even though in these sites the interaction is turned off. Note, though, the difference in scale from  Fig.~\ref{fig:mi2vsU}(a): The induced moment localization in free sites is almost seven times smaller than the moment acquired in repulsive sites for $U/t \approx 8$.  However, Fig.~\ref{fig:mi2vsU}(b) clearly shows that an increase in the ratio $L_{\rm U}/L_0$ increases $\langle \left(\hat m_\mathbf{i}^z\right)^2\rangle_0$ for finite values of $U/t$. It is interesting to note that for different superlattices with the same ratio $L_{\rm U}/L_0$, such as $L_{\rm U}=1$ $L_0=1$, and $L_{\rm U}=2$ $L_0=2$, the thinner non-interacting layer in the former favors the ``leakage" of local moment.
Ultimately, when $U \rightarrow\infty$, fermions on repulsive layers become completely localized  [see Fig.~\ref{fig:mi2vsU}(a)],
hopping between the free and repulsive sites is suppressed (see Sec. IV), pushing the local moment back to its non-interacting value on the free sites [Fig.~\ref{fig:mi2vsU}(b)].

To better understand the enhancement of moment localization in non-interacting sites, we probe the effects of the vicinity of a correlated layer, by considering separately the different lines that compose the non-interacting layer.  Figure~\ref{fig:si20}(a) shows the local moment in sites at the edges of the free layer (i.e., in the non-interacting line neighboring a repulsive layer; open symbols)  and 
along the central line of the layer (crossed symbols) -- see schematics in Fig.~\ref{fig:si20}(b) for $L_{\rm U}=1$ and $L_0=3$. When $L_0=1$ or $L_0=2$, edge and center lines coincide. Additionally, when $L_0 >2$, one can clearly see that the central line of the free layer is barely affected by the repulsive layers as  $\langle \left(\hat m_\mathbf{i}^z\right)^2\rangle_0$ remains very close to the non-interacting value (0.5). The increase in $\langle \left(\hat m_\mathbf{i}^z\right)^2\rangle_0$ is larger for superlattices with $L_0=1$; in these cases, each free line has two neighboring repulsive lines. The effect of the repulsive layers goes beyond nearest neighbors, as for fixed $L_0=1$, the local moment is larger for the superlatice with $L_{\rm U}=3$ than for the one with $L_{\rm U}=1$. 

\begin{figure}[!tb] 
  \includegraphics[width=0.98\columnwidth]{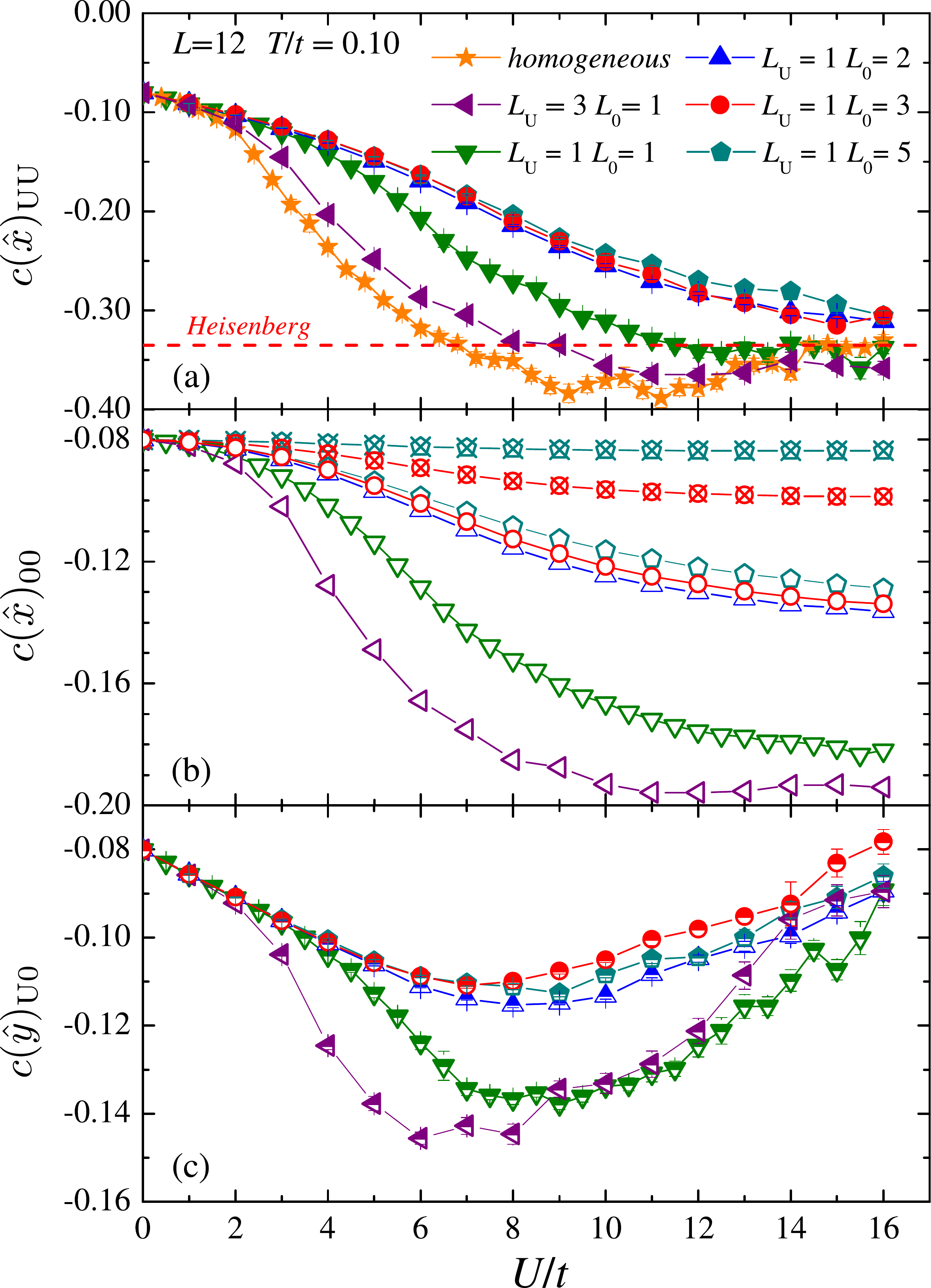}
 \caption{(Color online) Nearest-neighbor spin-spin correlation function as a function of $U/t$ at $T/t=0.10$ for $L=12$ lattices.  Correlations are calculated  along (a) repulsive (filled symbols) and (b) free (empty symbols) layers, and (c) perpendicular to the layers, between a free and a repulsive site (half-filled symbols).  
Stars represent the corresponding homogeneous system. The dashed line represent the Heisenberg model nearest-neighbor spin-spin correlations QMC result~\cite{sandvik}.}
 \label{fig:nnspin}
\end{figure}

We now turn to spin-spin correlation functions defined as $c(\mathbf{i}-\mathbf{j}) \equiv \langle \hat m_\mathbf{i}^z \hat m_\mathbf{j}^z\rangle$. Recently, single atom imaging for fermionic atoms trapped on optical lattices has been achieved in experiments with  $^6$Li~\cite{Parsons2015,Omran2015} and $^{40}$K atoms~\cite{Haller2015,Cheuk2015,Edge2015} enabling the measurement of spin-spin correlation functions in cold atom experiments~\cite{Cheuk1260,Boll2016,Parsons2016}.
Thus, we show in Fig.~\ref{fig:nnspin} the NN spin-spin correlation functions as a function of $U/t$ for different superlattices at $T/t=0.10$. 
The  negative values in all cases considered show  the antiferromagnetic arrangement. Similarly to what is seen for the local moment, nearest-neighbor (NN) spin correlations along repulsive sites [Fig.~\ref{fig:nnspin}(a)] approach the values for homogeneous systems as $L_{\rm U}/L_0$ increases. Note that for $L_0 >1$, increasing the width of the free layers has very little effect on the magnetic correlations along the repulsive sites. On the other hand, a ``leakage"  of magnetic correlations from the repulsive sites is present along free layers. This ``leakage" is strongly dependent on the neighboring sites. NN-spin-spin correlations along the central line of the free layer (crossed symbols) and also for the line at the edge of the  free layer (open symbols), are shown in Fig.~\ref{fig:nnspin}(b). For wide ($L_0=5$) free layers, NN-spin-spin correlations along  the central line of a non-interacting layer [crossed pentagons, Fig.~\ref{fig:nnspin}(b)] remain close to the non-interacting value. On the contrary,  correlations along the edges of the free layer  [open pentagons, and open circles, Fig. \ref{fig:nnspin}(b)] increase in modulus with $U/t$ and follow closely those for $L_{\rm U}=1$ and $L_0=2$ (open triangles). From this, one could think of a mechanism where at the ``interface" between the layers a hybridization of the orbitals of each site induces a singlet formation as a result of the strongly localized moment in the repulsive layer. It has been argued that this would occur in similar systems~\cite{Jiang12} where \textit{shielding} would prevent the correlations to spread inside the free layer.

However, there is no need to speculate. One can directly investigate the coupling of the adjacent spins in repulsive and free layers by probing the NN-spin-spin correlations along the $y$-direction, i.e., taken perpendicular to the direction of the layers. This is shown in Fig.~\ref{fig:nnspin}(c). Its dependence with the interaction $U$ is reminiscent to the effect of the local moment in free sites: For small values of $U$, the adjacent spins in repulsive and free layers display AF correlations that increase with increasing $U$,  reaching its maximum values for different SL's configurations for $U/t\approx8$. Larger interaction strengths reduce the magnitude of this coupling, due to the decrease in the local moment within the free sites as shown in Figs.~\ref{fig:mi2vsU}(b) and \ref{fig:si20}(a).
For $U/t\sim16$, the NN spin correlations are comparable to the non-interacting case,  denoting that the layers become uncorrelated. Hence we can rule out the shielding mechanism since even the local moments at the interface from the free layer side become less localized and the overall correlation with the repulsive layer is diminished for large $U$.

\begin{figure}[!tb] 
  \includegraphics[width=0.98\columnwidth]{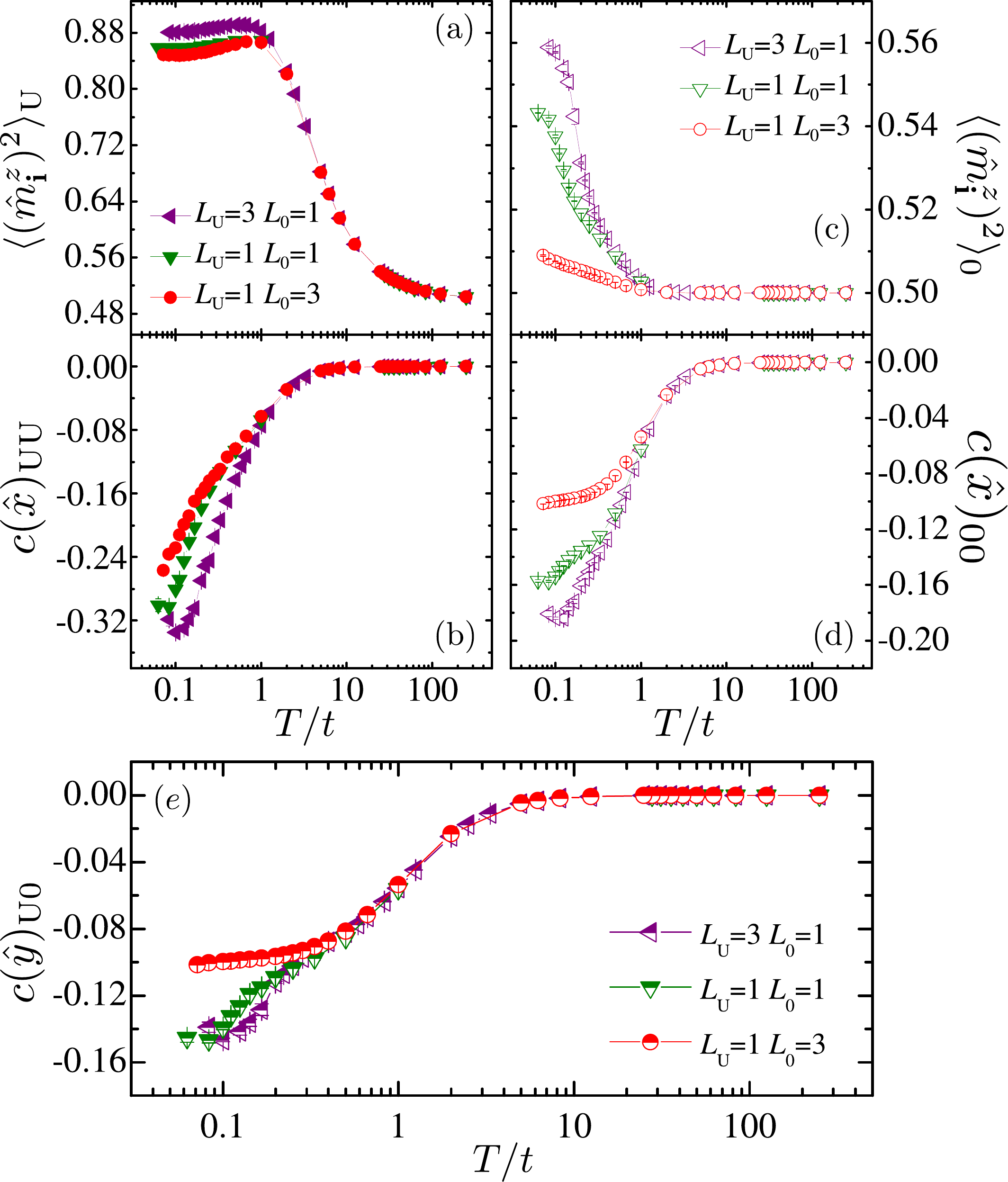}
 \caption{(Color online) (a) and (b) [(c) and (d)] display the local moment and NN spin-spin correlation function inside the repulsive [free] layer as a function of the temperature for three different SL structures. (e) also provides the temperature dependence but for the NN spin-spin correlation function where one site is at the repulsive layer and the other at the free one, computing, essentially, the coupling between the two types of layers.}
 \label{fig:corr_vs_T}
\end{figure}

Lastly, we want to understand how robust is this maximum spreading of the correlations that happens for $U/t\approx8$ from variations of the temperature. Figure~\ref{fig:corr_vs_T} displays the local moments and NN spin-spin correlation functions either inside a repulsive layer [panels (a) and (b)] or inside a free one [panels (c) and (d)], as a function of temperature. For large temperatures, all the quantities reduce to its uncorrelated value, i.e., the local moment is $\langle(\hat m_i^z)^2\rangle=0.5$ and the nearest neighbor correlation is vanishing. Note that the temperature used in the previous analysis ($T/t=0.1$) is already low enough to capture physics close to the ground state since most of the quantities are on the verge of saturation or already saturated. Hence we expect that in the limit $T\to0$, the decrease in moment localization for large values of interaction in sites within the free layer will be robust. Figure~\ref{fig:corr_vs_T}(e) also shows the NN spin-spin correlation function for sites at the interface between repulsive and free layers, where one can see that this coupling is larger, at low $T$, when the width of the free layer is small.

\subsection{Long-range ordering}
\begin{figure}[!tb] 
  \includegraphics[width=0.99\columnwidth]{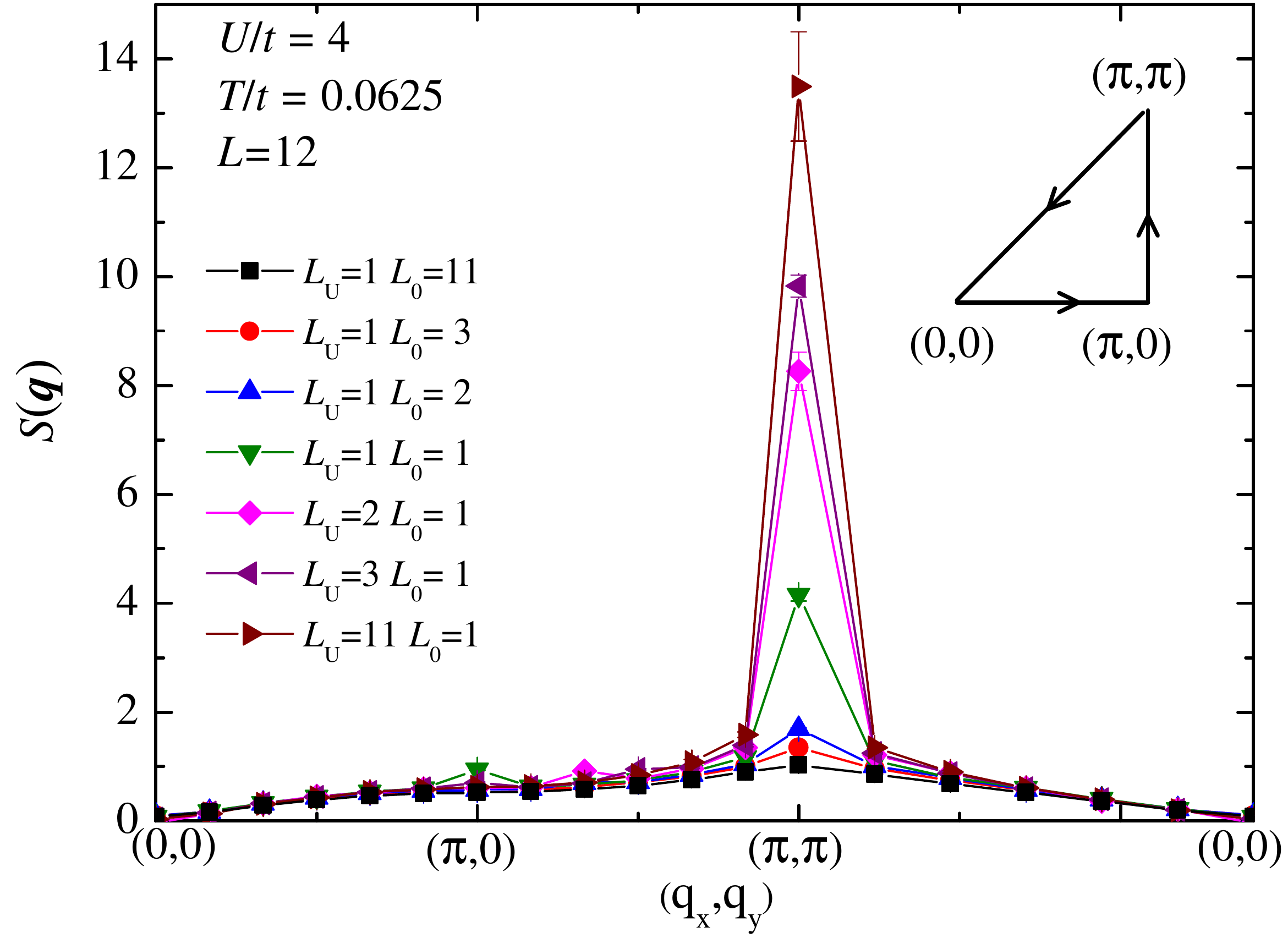}
 \caption{(Color online) Structure factor along a path in momentum space for different SL's with $L=12$, $U/t=4$ and $T/t=0.0625$.  All SL's shown display a dominant \textbf{q}=$(\pi,\pi)$ related with an overall AF ordering in both directions neglecting the underlying SL structure.}
 \label{fig:S_q}
\end{figure}

\begin{figure}[!tb] 
  \includegraphics[width=0.99\columnwidth]{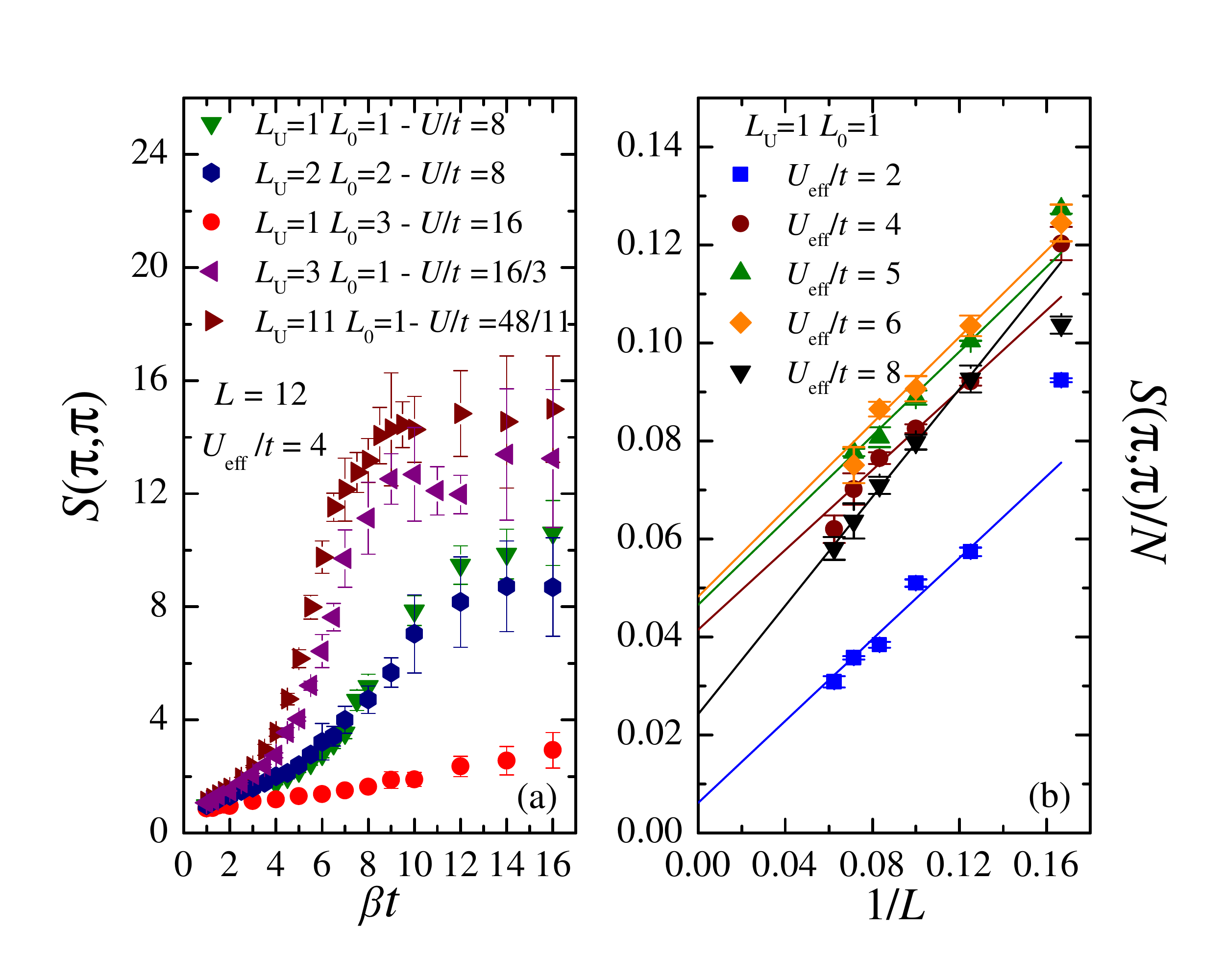}
 \caption{(Color online)  AF structure factor vs inverse temperature $\beta$ for different SL's with $L=12$ and $U_{\rm eff}/t =4$. It is clearly seen that an effective $U$ model does not explain all the features of magnetism in SL's. Finite-size dependence of AF structure factor for the SL's $L_{\rm U}=1$, $ L_0=1$ and various $U_{\rm eff}/t$. The linear extrapolation to limit $L\rightarrow\infty$ shows that the AF order is long ranged.}
 \label{fig:S_pi_pi}
\end{figure}

From a theoretical point of view, it is an open question whether selecting the interactions in a layered pattern still renders a global long-range magnetic order for the ground-state. The two-dimensional half-filled \textit{homogeneous} HM on the square lattice is known to display long-range AF order for any non-zero value of the local interaction energy $U$ at $T=0$~\cite{Hirsch85}. Given that for the SL´s this repulsive energy is not regularly distributed throughout the lattice, it is not obvious which magnetic arrangement minimizes the total energy when approaching this limit. In order to probe it, we calculate the magnetic structure factor
\begin{equation}
S(\mathbf{q})=\frac{1}{L^2}\sum_{\mathbf{i},\mathbf{j}}e^{i\mathbf{q}\cdot(\mathbf{i}-\mathbf{j})}c(\mathbf{i}-\mathbf{j});
\end{equation}
where \textbf{q} denotes the wave vector. Here, we make the choice of neglecting the periodicity of the SL and use as the wave vector the one associated with the homogeneous lattice. This will help to infer whether the long-range AF order is globally obtained regardless of the underlying superlattice structure. The  peaks in this quantity are related with the dominant spin ordering. For all the studied SL's, we observe a peak at \textbf{q}=$(\pi,\pi)$ related with AF ordering in both principal lattice directions, as shown in Fig. \ref{fig:S_q}. This peak becomes more pronounced as the number of repulsive sites is increased in relation to the number of free ones and one would be led to identify it with the increased average value of interaction strength. To characterize this picture, we can define an effective repulsive interaction,
\begin{equation}
U_{\rm eff} \equiv \frac{L_{\rm U}}{L_{\rm U}+L_0} U,
\end{equation}
and choose different SL's configurations, then setting a specific value of $U$ in order to keep the average repulsive interaction constant. Indeed, if magnetic properties were only ruled by $U_{\rm eff}$, different SL's would display the same $S(\pi,\pi)$ as long as  $U_{\rm eff}$ is kept fixed. However, this is not observed [Fig. \ref{fig:S_pi_pi}(a)] and the widths of the layers  strongly affect spin correlations. Not only the absolute values of the AF spin correlations are different when approaching the ground state, but also the temperature in which spin correlations reach their asymptotic value, which  occurs when the typical size of spin correlations $\xi$ becomes larger than linear lattice size $L$.

The presence of long ranged AF ordering in the thermodynamic limit is determined by a proper finite-size scaling analysis of the $\textbf{q}=(\pi,\pi)$ structure factor. Spin-wave theory~\cite{Huse88}, predicts that the finite-size corrections to $S(\pi,\pi)$ are linear in $1/L$:
\begin{equation}
\frac{S(\pi,\pi)}{N}=\frac{m_{\rm AF}^2}{3} + \frac{a}{L},
\end{equation}
where $m_{\rm AF}$ is the long-ranged AF order parameter and $a$ is a constant. This dependence is displayed in Fig.~\ref{fig:S_pi_pi}(b) for the $L_{\rm U}=1$ and $L_0=1$ SL showing that, indeed, long range AF order is present for all the analyzed values of $U_{\rm eff}$.

\begin{figure}[!tb] 
  \includegraphics[width=0.99\columnwidth]{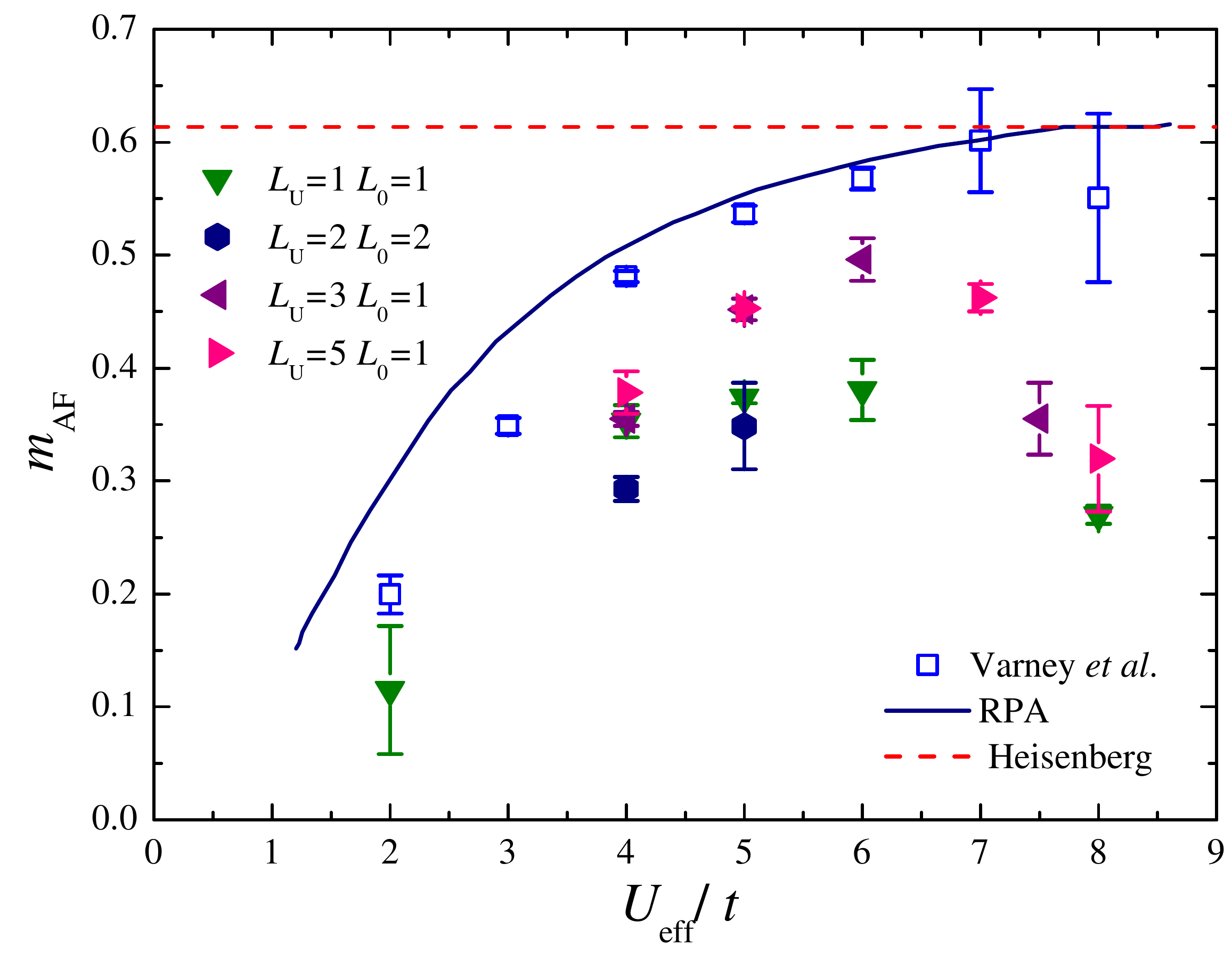}
 \caption{(Color online)  Staggered magnetization $m_{\rm AF}$ dependence with $U_{\rm eff}/t$ and four different SL's compared with homogeneous result obtained from Ref.\onlinecite{Varney09}. The continuous line is obtained from RPA approximation~\cite{Schrieffer89} and the dashed one is the QMC Heisenberg~\cite{sandvik} result.}
 \label{fig:m_AFvsU_eff}
\end{figure}

Compiling the values for the magnetic order parameter for different configurations, in Fig.~\ref{fig:m_AFvsU_eff} we compare $m_{\rm AF}$ for the superlattices with
 recent QMC data for the homogeneous lattice~\cite{Varney09}. The order parameter for different SL's is always smaller than for the homogeneous system, but  for  $U_{\rm eff}/t \lesssim 7$, it follows the same trend, i.~e., it increases with increasing interaction strength.  Moreover, the comparison among the different SL's configurations shows that this ordering depends non-trivially on the chosen pattern. Superlattices with the same $L_{\rm U}/L_0$ ratio, such as $L_{\rm U}=1$,  $L_0=1$ and $L_{\rm U}=2$,  $L_0=2$, do not always have the same value of $m_{\rm AF}$ when $U_{\rm eff}$ is kept fixed. Thus, an effective interaction mechanism is not sufficient to explain the observed long-range magnetic order. 
 For larger $U_{\rm eff}$, the order parameter does not saturate at the Heisenberg limit value (dashed line), as one would naively expect, and instead decreases. For large values of $U/t$, the free and repulsive layers decouple, as signaled by the reduced value of near-neighbor spin-spin correlations shown in Fig.~\ref{fig:nnspin}(c). In the $U \to \infty$ limit, the SL's become a set of uncoupled free and repulsive chains that are unable to sustain long range order in two-dimensions.

\section{Transport Properties}
\label{sec:transport}

\begin{figure}[!tb] 
  \includegraphics[width=0.9\columnwidth]{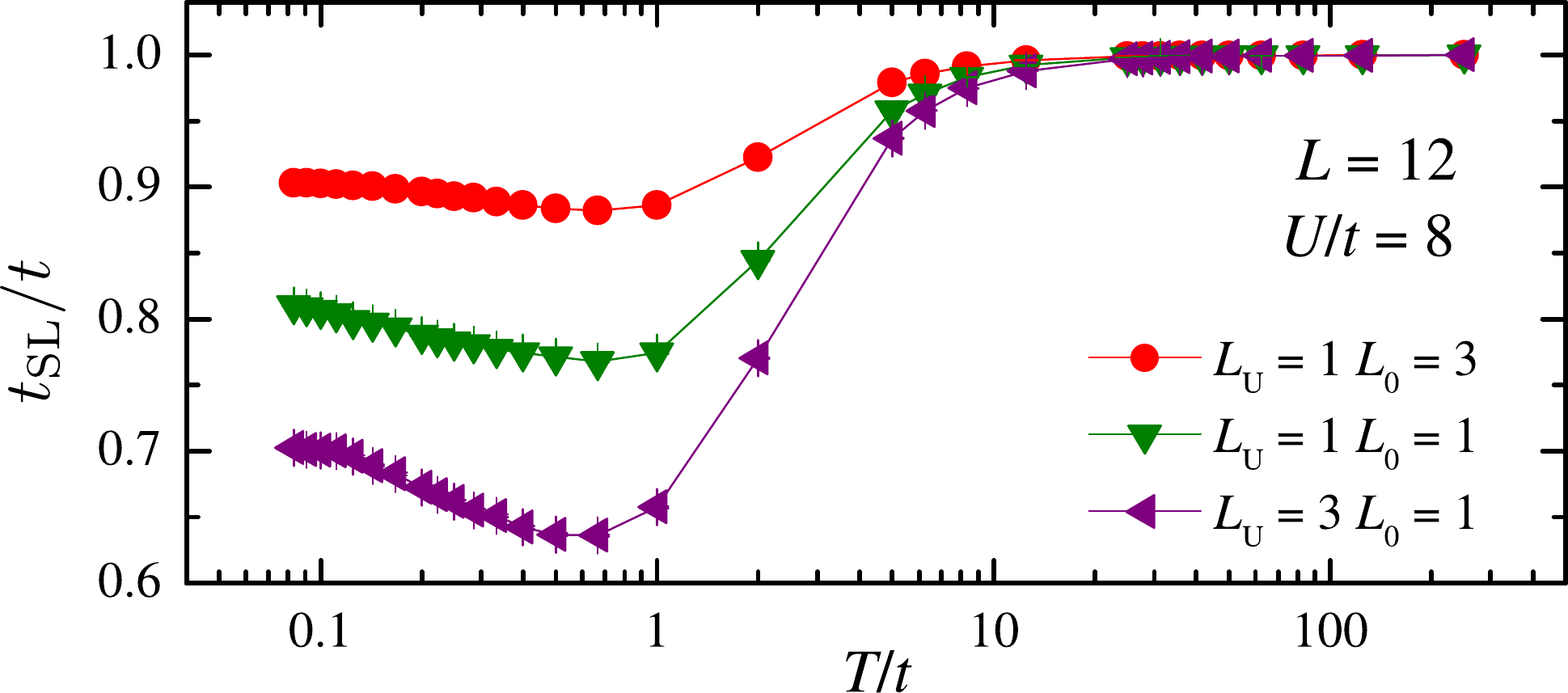}
 \caption{(Color online) Temperature dependence of the ratio of kinetic energies of superlattices to the non-interacting result. The lattice size is $12\times12$ and the interactions in the repulsive sites is $U/t=8$.}
 \label{fig:t_SL_t_vs_T}
\end{figure}

Better insight on the interplay of localization and delocalization in repulsive and free layers, initially obtained by investigating the spin correlations in the previous section, can be gained by checking some of the transport properties of the system. We start our study of the transport by analyzing the total effective hopping~\cite{White89b},
\begin{equation}
 \frac {t_{\rm SL}} {t} = \frac {\left\langle \sum_{\langle i,j\rangle,\sigma}
(\hat c^\dagger_{i\sigma} \hat c_{j \sigma}^{\phantom\dagger} +
\hat c^\dagger_{j\sigma} \hat c_{i \sigma}^{\phantom\dagger} )\right\rangle_{\rm SL}}{\left\langle \sum_{\langle i,j\rangle,\sigma}
(\hat c^\dagger_{i\sigma} \hat c_{j \sigma}^{\phantom\dagger} +
\hat c^\dagger_{j\sigma} \hat c_{i \sigma}^{\phantom\dagger} )\right\rangle_{0}},
\end{equation}
which we define as the ratio of the kinetic energy on a superlattice, averaged over both the directions, along and across the layers, to its non-interacting counterpart value. We start by checking the temperature dependence of this quantity in Fig.~\ref{fig:t_SL_t_vs_T} for different SL structures at $U/t=8$. In the high-temperature limit ($T\gg t$), the effect of interactions (either layered or homogeneously distributed) is negligible and the kinetic energy is essentially equivalent to the kinetic energy of the non-interacting system ($t_{\rm SL}/t\to1$). For decreasing temperatures, the actual pattern of interactions affects the overall charge mobility and  the correspondent kinetic energy for interacting SL's drops to a fraction of the non-interacting value which is inversely proportional to $L_{\rm U}/L_0$. In the following, we will focus on values of temperature $T/t=0.1$, which is small enough to capture the physical aspects when approaching the ground state for the different superlattices structures, since the kinetic energy is either converged or in the verge of convergence for decreasing temperatures, but not so small to render unnecessarily complicated large simulations. It is important to notice that quantum fluctuations are responsible for the fact that the kinetic energy is still finite when approaching the zero temperature limit. 

\begin{figure}[!tb] 
  \includegraphics[width=0.99\columnwidth]{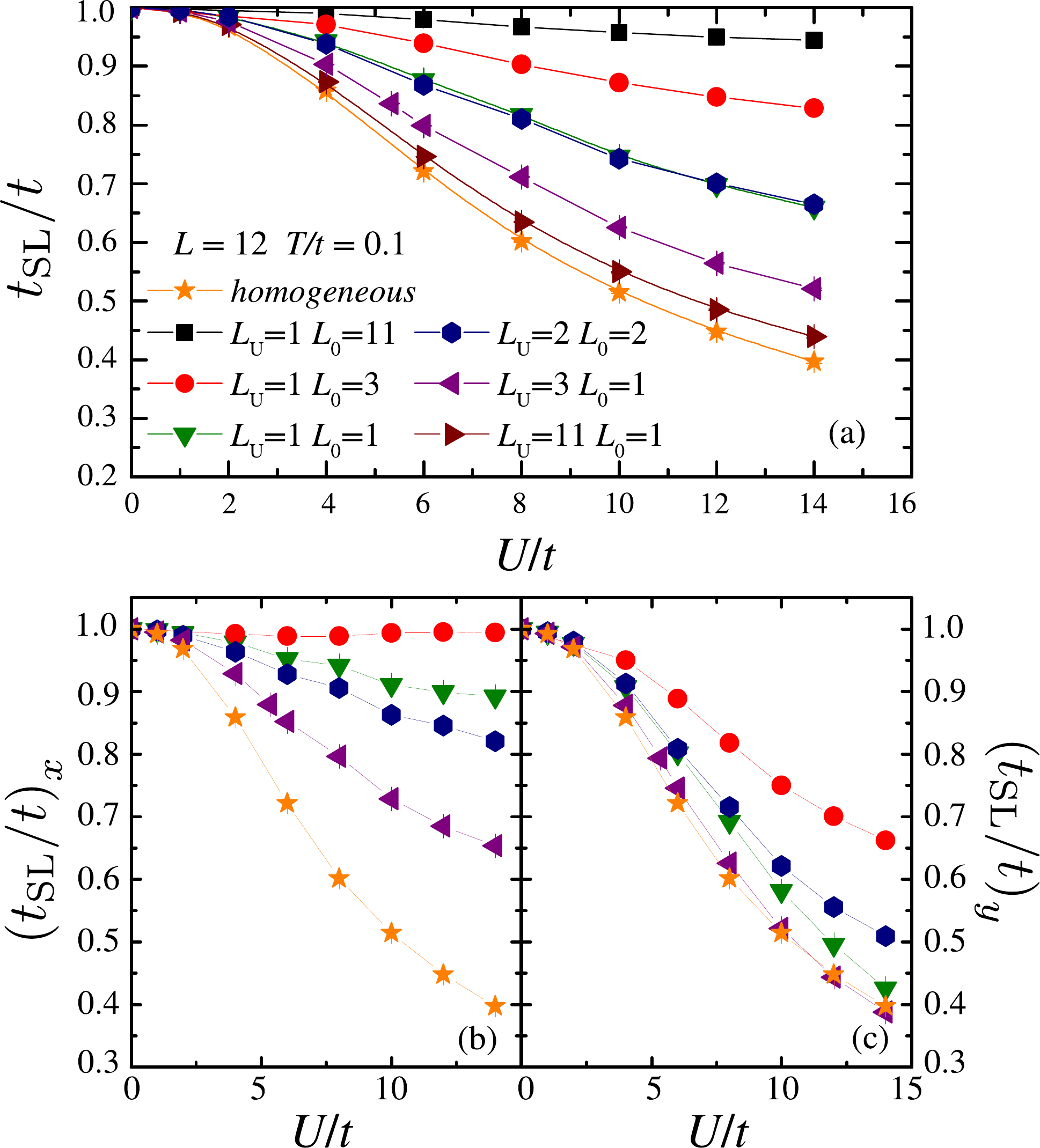}
 \caption{(Color online) (a) Effective hopping as a function of interaction strength  $U/t$, for the homogeneous system and different  superlattices  with  $ L=12 $ and  $T/t= 0.10$. (b) and (c) with the effective hopping contribution split along direction along and across layers, respectively.}
 \label{fig:t_effvsU_and_x_y}
\end{figure}

Figure~\ref{fig:t_effvsU_and_x_y}(a) shows the $U$-dependence of $t_{\rm SL}/t$  for different superlattices and also for the homogeneous system for $12\times12$ lattices at $T/t=0.1$. In all cases, increasing $U$ induces charge localization in at least the repulsive layers and, therefore, decreases the total hopping energy in comparison to the non-interacting limit. We can see that $t_{\rm SL}/t$ is strongly dependent on the ratio $L_{\rm U}/L_0$, converging towards the non-interacting limit ($t_{\rm eff}/t=1$) as  $L_{\rm U}/L_0 \to 0 $ (see, for instance, black squares for $L_{\rm U}=1$ and $L_0=11$) and approaching the homogeneous system results as $L_{\rm U}/L_0$ increases (see right triangles for $L_{\rm U}=11$ and $L_0=1$).

The large anisotropic character introduced by the layered construction makes it important to analyze the effective hopping along ($x$) and across ($y$) the direction of the layers, which we define as:
\begin{equation}
\left(\frac {t_{\rm SL}} {t}\right)_{\alpha}=
\frac {\langle  \sum_{j,\sigma} \hat c^\dagger_{j+\hat{\alpha}\sigma} \hat c_{j \sigma}^{\phantom\dagger}+ \hat c^\dagger_{j\sigma} \hat c_{j +\hat{\alpha}\sigma}^{\phantom\dagger}  \rangle_{\rm SL}}  {\langle \sum_{j,\sigma} \hat c^\dagger_{j+\hat{\alpha}\sigma} \hat c_{j \sigma}^{\phantom\dagger}+ \hat c^\dagger_{j\sigma} \hat c_{j +\hat{\alpha}\sigma}^{\phantom\dagger}  \rangle_0},
\end{equation}
where $\hat \alpha = \hat x$ or $\hat y$.

\begin{figure}[!tb] 
  \includegraphics[width=0.99\columnwidth]{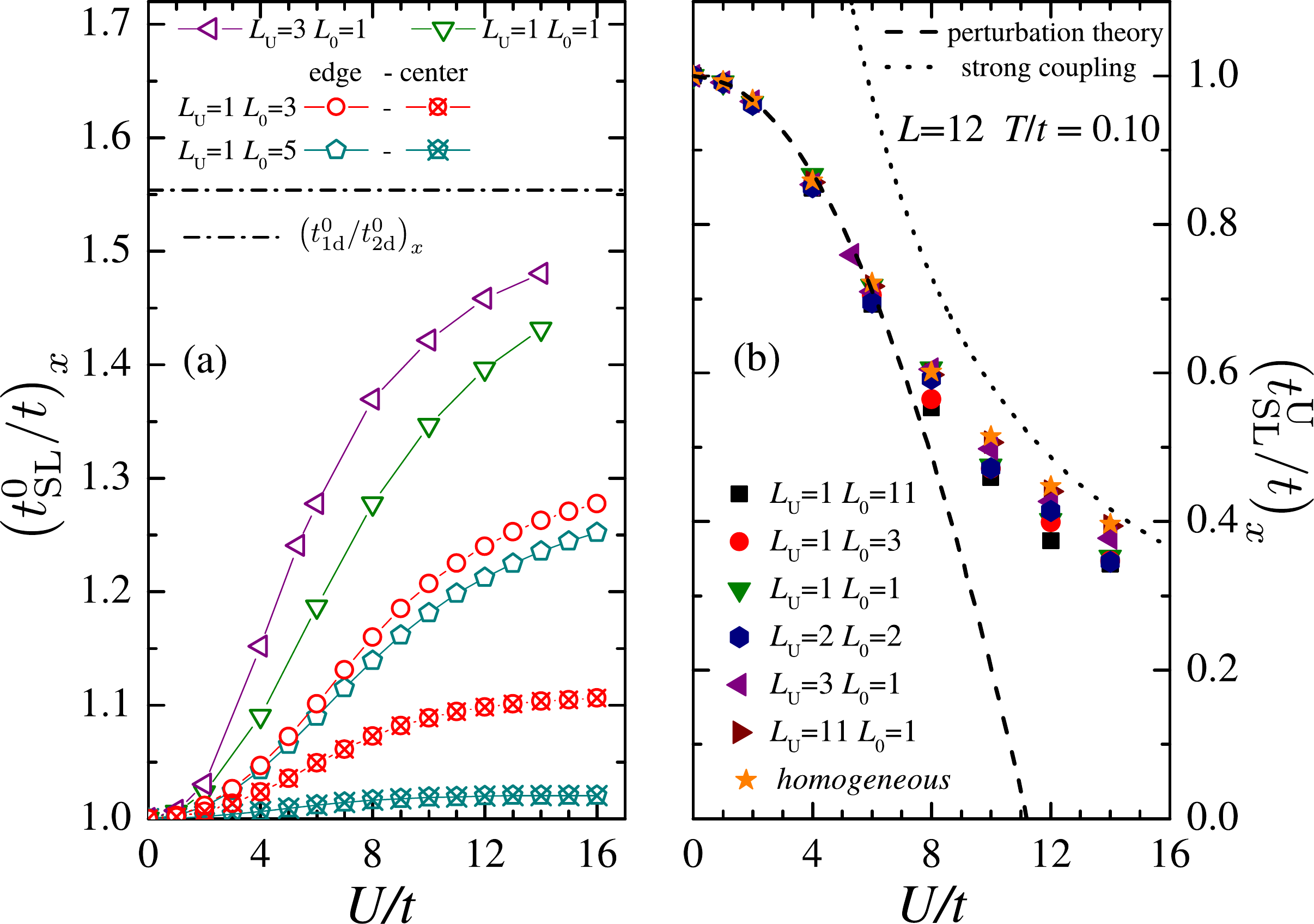}
 \caption{(Color online) Ratio between the $x$-component of the kinetic energy at a finite $U/t$ to the $U=0$ case for $L=12$ lattice with $T/t=0.10$ for different SL´s. In (a) we plot the hopping contribution from the free layers and in (b) the same for the repulsive ones. The filled (empty) symbols denote electronic transport between repulsive (free) sites. To avoid misleading interpretation due to the different number of repulsive ($N_{\rm U}$) and free ($N_{{\rm U}=0}$) sites, in this case we have normalized the results by the number of sites of each type. In (b) we also include the analytical results in the extreme limits of $U/t\ll1$ (perturbation theory) and $U/t\gg1$ (strong coupling)~\cite{White89b} and the contribution to the kinetic energy between repulsive sites smoothly interpolates between both limits albeit the layered distributed interactions.}
 \label{fig:txU_0_vsU}
\end{figure}

At first sight, one would expect that the anisotropy favors the electronic transport along the direction of the layers. This is indeed the case, as can be readily observed in Figs.~\ref{fig:t_effvsU_and_x_y}(b) and ~\ref{fig:t_effvsU_and_x_y}(c), where the repulsive interaction splits the two contributions of the effective hopping. In fact, the largest contribution to the transport in the direction parallel to the layers should be related to stripes formed by free sites since  within  the repulsive layers  local moment formation is favored [see Fig.~\ref{fig:mi2vsU}(a)] and, consequently, the  mobility is reduced. We separate the contribution of the kinetic energy along the layers between  the repulsive and free layers via,
\begin{equation}
\left(\frac {t^{\rm U,0}_{{\rm SL}}} {t}\right)_x=\frac {   \langle  \sum_{j,\sigma} \hat c^\dagger_{j+\hat{x}\sigma} \hat c_{j \sigma}^{\phantom\dagger}+ \hat c^\dagger_{j\sigma} \hat c_{j +\hat{x}\sigma}^{\phantom\dagger}  \rangle^{\rm U,0}_{\rm SL}}  {\langle  \sum_{j,\sigma}  \hat c^\dagger_{j+\hat{x}\sigma} \hat c_{j \sigma}^{\phantom\dagger}+ \hat c^\dagger_{j\sigma} \hat c_{j +\hat{x}\sigma}^{\phantom\dagger}  \rangle_0},
\end{equation}
where the denominator refers to the average kinetic energy along one direction of a two-dimensional non-interacting square lattice.
Figure~\ref{fig:txU_0_vsU} shows how the separate contribution of the hopping depends on the strength of interactions in free [panel (a)] and repulsive layers [panel (b)]. One observes, in the latter, that the transport along repulsive sites does not significantly change for different SL configurations, remaining close to the correspondent homogeneous results and smoothly interpolating the limits of small interactions, obtained within perturbation theory (dashed curve), and the strong coupling limit~\cite{White89b} (dotted line). On the other hand,  in the former, as $U/t$ is increased, the contribution to the kinetic energy due to the hopping between free sites is always larger than one, i.e., enhanced in comparison to the contribution to the kinetic along one direction in a completely non-interacting two-dimensional lattice. In this case, when comparing  different SL's patterns we can see that the enhancement is maximum when the free layer is  just one-site thick ($L_0=1$) and increases with increasing $L_{\rm U}/L_0$. This feature is an indicative of the change of dimensionality, as a result of increased interactions, being related to the picture where free layers become uncoupled to the repulsive ones as the large $U/t$ limit is approached, which was also inferred when investigating the magnetism in Sec.~\ref{sec:magnetism}. This scenario is supported by noting that the  
the kinetic energy along the free layers of the SL systematically converges to the kinetic energy of a one-dimensional non-interacting chain (dashed-dotted line in Figure~\ref{fig:txU_0_vsU}(a)) for $U/t\gg1$. 

One can also get useful physical information by 
investigating wider free layers as in the SL with configurations $L_{\rm U} = 1$ and $L_0=3$ or 5 [circles and pentagons, respectively, in Fig.~\ref{fig:txU_0_vsU}(a)]. When analyzing the contribution of the hopping in the free layers, we see that, the wider the free layer is, the less its center is affected by the repulsive sites.  Still, for the free sites at the edge between the two regions, the enhancement of kinetic energy along the $x$ direction is substantial, reaching $\sim 25\%$ of increase for $U/t=16$.

It is still an open question to examine other quantities that could potentially fully characterize transport, and definitively quantify whether the superlattice displays a metal-insulator transition when approaching $T=0$ at large $U/t$ limit. Among them one could highlight the dc-conductivity~\cite{Randeria1992,Trivedi1996} and the Drude weight~\cite{Scalapino93,Mondaini12} both of which can be computed using imaginary time-dependent correlation functions in QMC calculations. However, while the kinetic energy per site does not change substantially when using different system sizes, we have checked that the Drude weight and the dc-conductivity possess dramatic finite size effects. Besides, as we argued before, the fact that some of the superlattices are not commensurate with the system size makes a proper finite-size scaling analysis elusive. Future studies may shed light on this issue and unequivocally answer the question of whether the increase of repulsive interactions may induce a Mott-insulator to anisotropic metal transition in the large $U$ limit.


\section{Temperature scales}

We now turn to the study of the temperature scales that characterize the superlattices. The Mermin-Wagner theorem~\cite{merminwagner} establishes that long range order is only possible at $T=0$ for two-dimensional systems with continuous symmetry. Nonetheless, one can define finite temperature scales where strong magnetic and charge correlations start to develop.  The knowledge of such temperature scales is relevant in the context of fermionic cold atom experiments as spin and charge correlations in two-dimensional systems were recently measured. \cite{Cheuk1260,Parsons2016}

\label{sec:tdynprop}
\subsection{Spin susceptibility}
\begin{figure}[!tb] 
  \includegraphics[width=0.9\columnwidth]{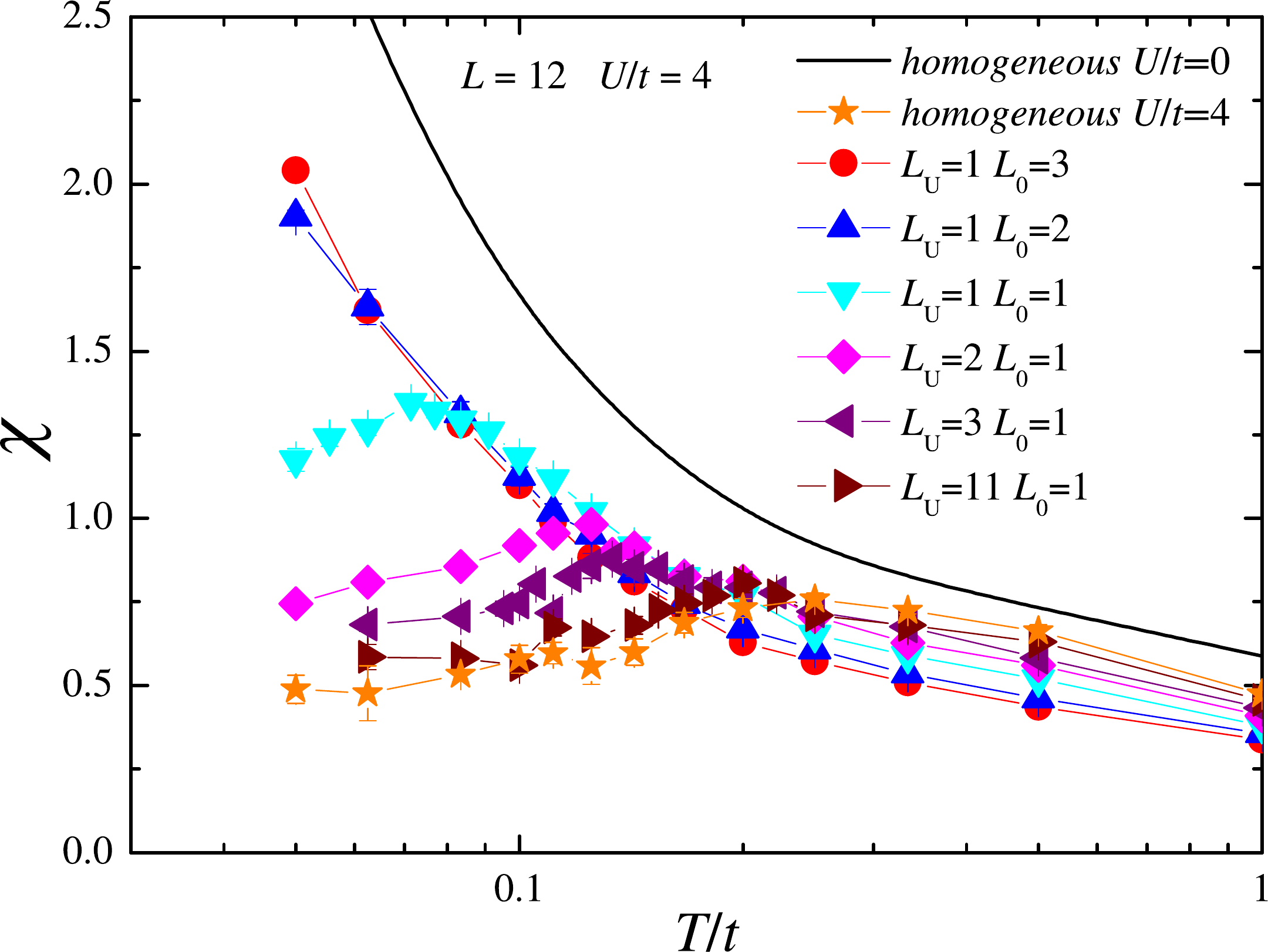}
 \caption{(Color online) Uniform spin susceptibility $\chi$ as a function of temperature for various SL's with $U/t=4$ and for homogeneous systems with $U/t=0$ (line) and $U/t=4$ (stars) in $L=12$ lattices. The peak in this quantity defines $T_{\rm spin}$, the onset of antiferrogmagnetic fluctuations.}
 \label{fig:chi_T}
\end{figure}

A crossover temperature $T_{\rm spin}$, below which spin correlations grow rapidly, can be obtained by the temperature where the uniform magnetic susceptibility $\chi(\textbf{q}=0,T)=\beta S(\textbf{q}=0)$ peaks~\cite{tspin}. Figure \ref{fig:chi_T} shows the susceptibility as a function of $T$ for different superlattices at $U/t=4$. For the SL's with $L_{\rm U}/L_0<1$ the crossover temperature is below $T=t/20$, the smaller temperature reached in most of our simulations and far beyond what can be reached under cold atoms experiments.  The \textit{finite} lattices are AF ordered at non-zero temperatures and we can associate $T_{\rm spin}$ with the finite N\'eel temperature for 2D lattices calculated within the random phase approximation (RPA) and Hartree-Fock calculations: $T_{\rm N} \propto t \exp[-2\pi \sqrt{t/U}]$~\cite{Hirsch85}. Figure~\ref{fig:Tlow-Tspin} compiles the positions of the peaks, $T_{\rm spin}$ (empty symbols), as a function of $U_{\rm eff}$ for different superlattice patterns, different system sizes and $U/t=4$, together with the RPA form. It clearly shows that the crossover temperature is governed by the effective interaction strength, essentially being independent of the underlying superlattice structure.

\subsection{Specific heat}
\begin{figure}[!tb] 
  \includegraphics[width=0.9\columnwidth]{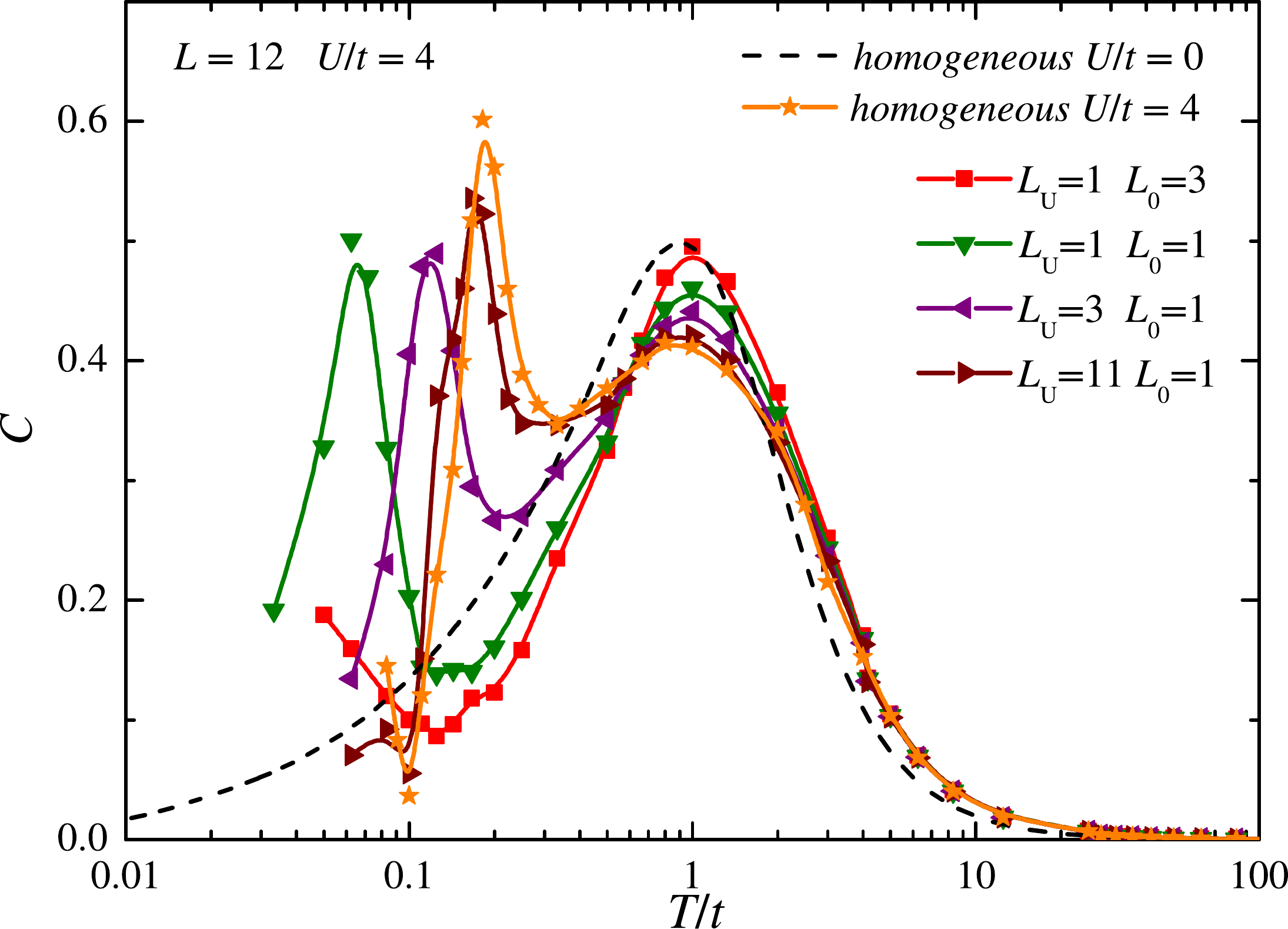}
 \caption{(Color online) Temperature dependence of specific heat for four different SL configurations with $L=12$ and the corresponding homogeneous cases with $U/t=0$ and $U/t=4$. While the high-temperature peak position is roughly constant at $T_{\rm high}\approx t$, the low-temperature one significantly varies for the SL's presented and displays increasing $T_{\rm low}$ as the $U_{\rm eff}$ becomes higher and closer to the homogeneous $U/t=4$ case. Lines are guides to the eye.}
 \label{fig:C_T}
\end{figure}

\begin{figure}[!tb] 
  \includegraphics[width=0.99\columnwidth]{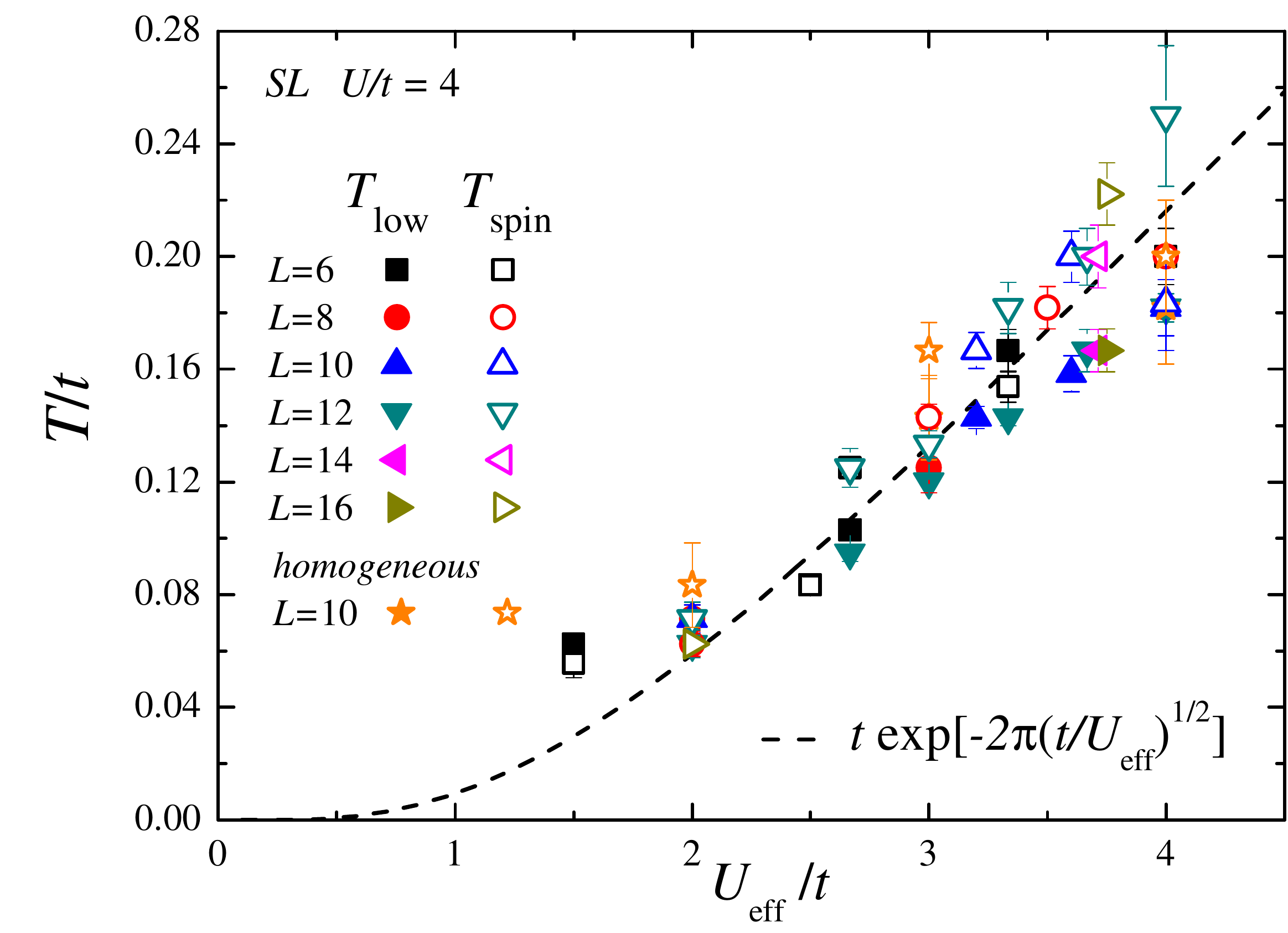}
 \caption{(Color online) Position of the low-$T$ peak of the specific heat (full symbols) and peak of the spin-susceptibility (open symbols) as a function of $U_{\rm eff}/t$ for different SL's with $U/t=4$ together with the homogeneous lattice results. Dashed line corresponds to an RPA-like form for the temperature scale in which antiferromagnetic fluctuations occur: $T \propto t \exp[-2\pi \sqrt{t/U_{\rm eff}}]$.
 }
 \label{fig:Tlow-Tspin}
\end{figure}

Another quantity that can provide insight into the temperature scales of the system is the specific heat $C$. We use the definition $C(T)=dE/dT$ to obtain the specific heat  by numerical differentiation of the total energy $E(T)$. Figure~\ref{fig:C_T} shows $C(T)$ for different SL's with $L=12$ at $U/t=4$ and also the results for the homogeneous case in the non-interacting (dashed line) and interacting limits. In the latter, the specific heat is known to display a two peak structure~\cite{Paiva01,Duffy97}: a broad high-temperature peak at $T_{\rm high}$, associated  with ``charge fluctuations,''  and a sharp peak at $T_{\rm low}$  associated with ``spinfluctuations.'' These denote temperature below which these degrees of freedom start to freeze.  Note that, for fixed $U$,  while the high$-T$ peak position is very similar for all superlattices, $T_{\rm low}$ strongly depends on the superlattice pattern, shifting to lower temperatures as the ratio $L_{\rm U}/L_0$ is reduced. We were not able to resolve the peak when $T_{\rm low} < t/20$, which is the case when $L_{\rm U}/L_0<1$. For the other cases, Fig.~\ref{fig:Tlow-Tspin} shows the dependence of $T_{\rm low}$ (filled symbols) with $U_{\rm eff}$, for different system sizes and different superlattice patterns with $U/t=4$.  Similar to $T_{\rm spin}$,  $T_{\rm low}$ also defines a temperature scale where AF correlations become relevant, therefore, $T_{\rm low}$ obeys the same  RPA-like form for small values of $U_{\rm eff}$.

\begin{figure}[!tb] 
  \includegraphics[width=0.99\columnwidth]{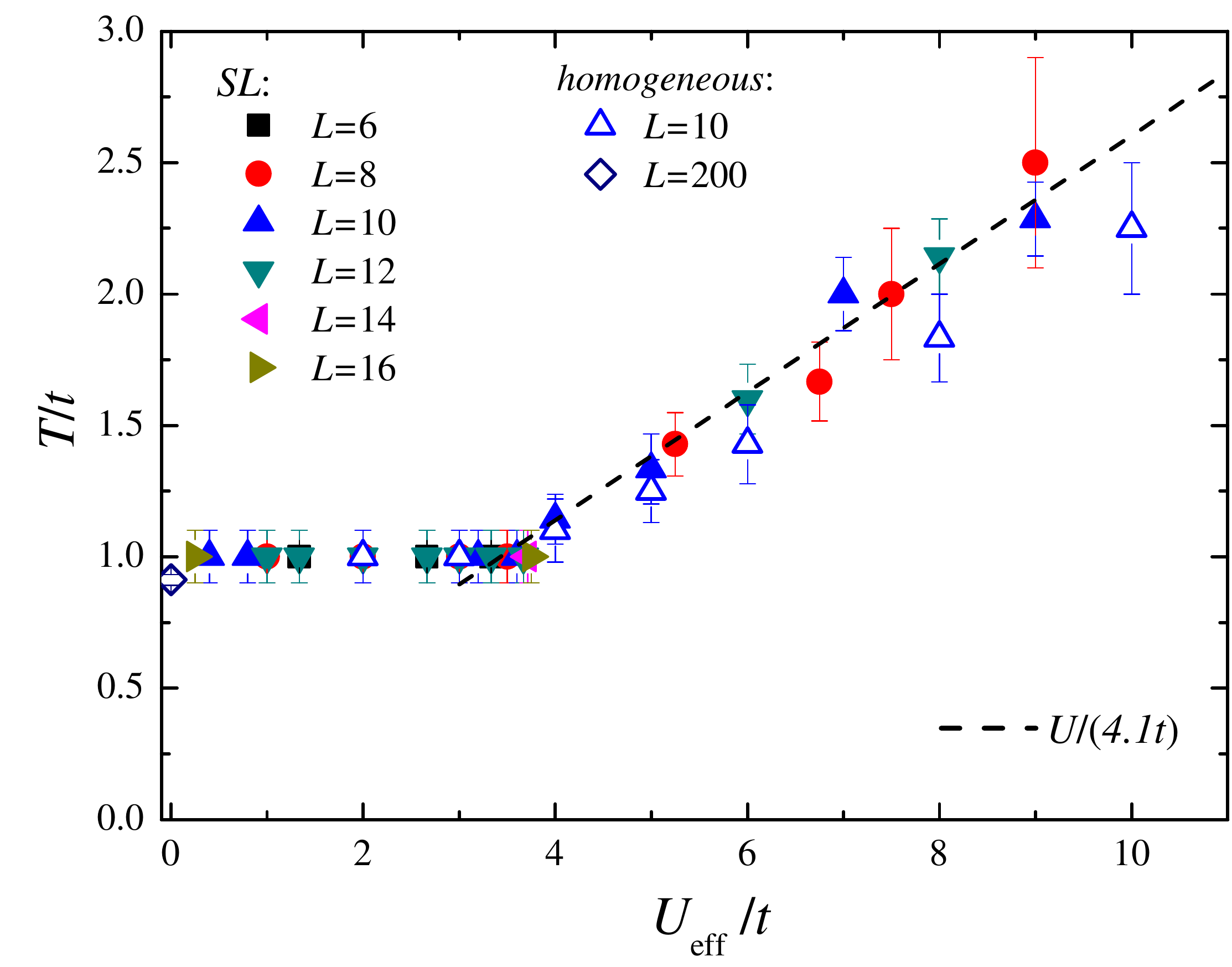}
 \caption{(Color online) Position of the high-$T$ peak of specific heat for different SL's as a function of $U_{\rm eff}$, the dashed line denotes the linear extrapolation to the data presented in the strong coupling limit ($T\sim U/4.1$). Error bars denote the confidence interval of estimating the peak from the data of the specific heat}
 \label{fig:T_high}
\end{figure}

To better understand the position of the peaks and its dependence with the interaction strength, it is instructive to recall that the energy can be separated in kinetic and potential parts which separately contribute to the specific heat. In a strong coupling picture, the high-$T$ peak can be understood as the temperature where double occupations ($\langle \hat d\rangle $) are suppressed and, therefore, is governed by the contribution of the potential energy $P= U \langle \hat d \rangle$ to the specific heat.  In this regime, $T_{\rm high}$ has a linear dependence with $U_{\rm eff}$ as $T_{\rm high}\approx U_{\rm eff}/4.1$, as can be seen in  Fig.~\ref{fig:T_high}. In contrast, for weak interactions, the high-$T$ peak is determined by the kinetic energy contribution to the specific heat~\cite{Paiva01}. Figure~\ref{fig:C_T} clearly shows that  $T_{\rm high}$ closely follows the non-interacting peak: $T_{\rm high}\approx t=1$, for all superlattices shown at $U/t=4$. 

Although in the previous section we observed that an effective interaction cannot explain the discrepancies in spin-spin correlation functions or the ground state values of the order parameter for different SL's, the temperature scales presented in this section are clearly ruled by $U_{\rm eff}$.

\subsection{Entropy}

The entropy is a central quantity for cold atoms, as it can be obtained more easily on  experiments than the temperature. Understanding the behavior of the entropy as a function of the temperature for  different interaction strengths and SL's configurations can help in devising new cooling schemes, which are useful if one is concerned with the emulation of the low-temperature physics of strongly correlated systems. Here, we obtain the entropy per particle in units of the Boltzmann constant $k_{\rm B}$, by integrating the energy per particle $e\equiv E/N$ in inverse temperature $\beta$:

\begin{equation}
\frac {S(\beta)}{Nk_{\rm B}}= \rm{ln}(4)+ \beta e(\beta) - \int_0^\beta d\beta^\prime  e(\beta^\prime).
\end{equation}

\begin{figure}[!tb] 
  \includegraphics[width=0.9\columnwidth]{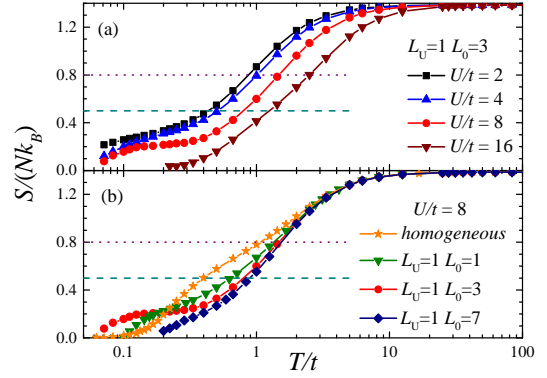}
 \caption{(Color online) Entropy as a function of temperature for fixed $L_{\rm U}=1$ and $L_0=3$ and $U/t=2$, 4, 8 and 16 (a) and for fixed $U/t=8$, and different SL configurations in a lattice with $L=8$. Dotted (dashed) lines represent adiabats with $S/(Nk_B)=0.8$ (0.5).}
 \label{fig:entropy}
\end{figure}

Figure~\ref{fig:entropy}(a) shows this quantity as a function of the temperature for the SL with $L_{\rm U}=1$ and $L_0=3$ and different values of the interaction strength $U/t$. For a fixed entropy  value  (see dotted and dashed horizontal lines), increasing $U/t$, which can be tuned by adjusting Feshbach resonances~\cite{Bloch08}, will lead to heating of the system, i.e., the temperature increases with increasing repulsive interactions. Figure~\ref{fig:entropy}(b), on the other hand, displays the temperature dependence of $S/(Nk_B)$ for fixed  $U/t=8$ and different SL configurations, as well as for the homogeneous system with the same interaction value.  If we start with an SL with $L_{\rm U}=1$ and $L_0=7$ (diamonds), and adiabatically turn the interaction on within lines of sites on the free layers, changing the pattern to a different SL configuration with $L_{\rm U}=1$ and $L_0=3$  (circles), and then $L_{\rm U}=1$ and $L_0=1$ (down triangles), and finally reaching the homogeneous system (stars), there is a range of entropies $0.2 \lesssim S/Nk_{\rm B} \lesssim 1.0$ for which the temperature is reduced.  

\begin{figure}[!tb] 
  \includegraphics[width=0.99\columnwidth]{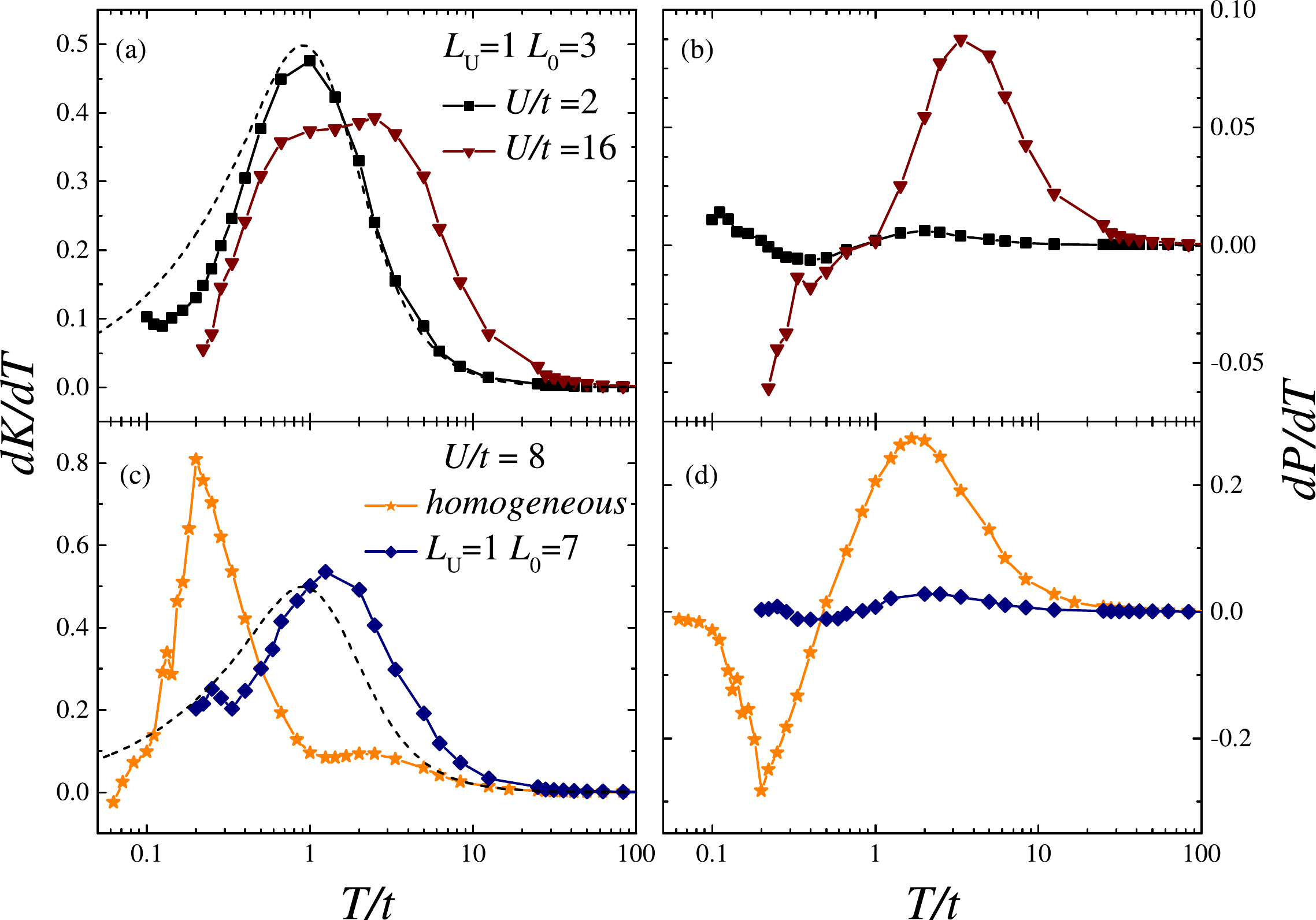}
 \caption{(Color online) Panels (a) and (c) [(b) and (d)] show the temperature dependence of the kinetic [potential] energy contributions to the specific heat for a $L=8$ lattice. Panels (a) and (b) focus in a given SL with configuration $L_{\rm U} = 1$ and $L_0=3$ and different interactions, while (c) and (d) compare the homogeneous result with a SL ($L_{\rm U}=1$ and $L_0=7$) for a given $U/t=8$. Dashed lines depict the non-interacting result.}
 \label{fig:KeP}
\end{figure}

In both panels of Fig.~\ref{fig:entropy}, $U_{\rm eff}/t$ is increased; in (a) by increasing the interaction $U/t$ while keeping $L_{\rm U}/L_0$ fixed, while in (b), by increasing $L_{\rm U}/L_0$ and keeping $U/t$ fixed.  To undestand how the former leads to heating and the latter to cooling (in an intermediate range of temperatures) let us examine separately the contributions of the kinetic ($K$) [Figs.~\ref{fig:KeP}(a) and ~\ref{fig:KeP}(c)] and potential ($P$) [Figs.~\ref{fig:KeP}(b) and ~\ref{fig:KeP}(d)] energies to the entropy. It is useful to remember that $S = \int_0^T dT^\prime C/T^\prime $, where $C=dE/dT$ is the specific heat and $E=K+P$. For fixed $L_{\rm U}=1$, $L_0=3$, and small $U/t$, the system is similar to the non-interacting one, with most of the entropy coming from the kinetic energy contribution. In the opposite limit (large $U/t$), the contribution from the kinetic energy moves to higher temperatures and the one associated with the potential energy becomes more relevant. The potential energy contribution comes from the double occupancies $\langle \hat d\rangle$ on repulsive sites. The double occupancies are directly related to the local moments via $\langle \hat d\rangle = [\langle\left( \hat n_{i\uparrow} + \hat n_{i\downarrow}\right)-(\hat m_{i}^z)^2\rangle]/2$. Starting at high temperatures, where it assumes its uncorrelated value $\langle \hat d\rangle=1/4$ for all $U/t$, as $T/t$ decreases, $\langle \hat d\rangle$ also decreases. For small $U/t$, $\langle \hat d\rangle$ hardly changes with $T/t$ and the contribution of $dP/dT$ is small. For large $U/t$, on the other hand, the change in  $\langle \hat d\rangle$ gives rise to the high-$T$ peak in $dP/dT$ (and also in $C$, see Figs.~\ref{fig:C_T} and \ref{fig:T_high}). The freezing of the charge degrees of freedom as $U/t$ increases at the repulsive sites leads to heating, observed in Fig.~\ref{fig:entropy}(a). 

For fixed $U/t$ and increasing $L_{\rm U}/L_0$, the behavior is similar to the previous one  at smaller $U_{\rm eff}/t$: most of the contribution to the entropy comes from the kinetic energy at temperature scales around $T/t \sim 1$.
As $L_{\rm U}/L_0$ increases, two effects take place, first  a high-T peak also develops in the contribution from the potential energy, similar to what is seen in the previous case, and, second, the peak in the kinetic energy derivative becomes sharper and moves to lower temperatures. This second effect is the responsible for cooling the system as $L_{\rm U}/L_0$ is increased.

\begin{figure}[!tb] 
  \includegraphics[width=0.99\columnwidth]{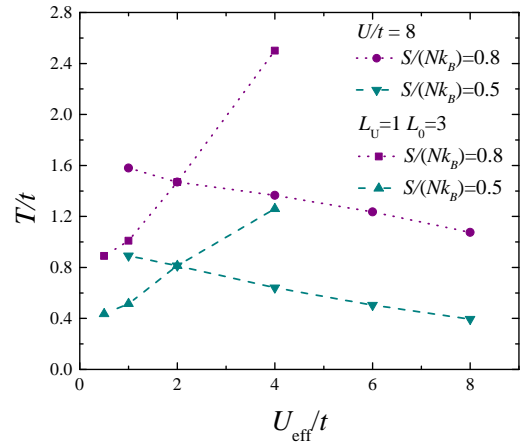}
 \caption{(Color online) Temperature as a function of $U_{\rm eff}/t$ for fixed $S/(Nk_{\rm B})=0.5$ (dashed lines) and 0.8 (dotted lines), for fixed $L_{\rm U}=1$ $L_0=3$ (squares and up triangles) and for fixed $U/t=8$ (circles and down triangles).}
 \label{fig:adiabatical}
\end{figure}

Figure~\ref{fig:adiabatical} shows how the temperature changes with $U_{\rm eff}/t$ along the adiabats with $S/(Nk_B)=0.5$ and 0.8. For a fixed SL, increasing $U_{\rm eff}/t$ leads to heating. This effect is more pronounced than in the homogeneous case~\cite{tspin}. Keeping the SL fixed and increasing $U/t$ increases the temperature by a factor of three at $S/(Nk_B)=0.5$ for $U_{\rm eff}/t$ going from 0.5
to 4.  Conversely,  if there is a way to experimentally turn on the interaction strength adiabatically on sites from the free layers, this could be a useful way to cool down the system and achieve lower temperatures in comparison to homogeneous ones.  Starting from a system with $L_{\rm U}=1$,  $L_0=7$ with $U/t=8$ and $T/t=0.89$, turning on the interactions at the free layers until the homogeneous system is achieved, leads to a final temperature of $T/t\simeq0.40$, more than a factor of two below the initial one.

\section{Conclusions}
\label{sec:concl}
In summary, we have employed quantum Monte Carlo methods to perform a thorough analysis of the half-filled Hubbard model on a two-dimensional lattice with layer distributed onsite interactions $U$ to understand how it affects the magnetism and charge dynamics. We have found that although the superlattices contain layers with sites possessing vanishing interactions, they are still able to sustain a global antiferrogmagnetic long-range order at finite values of the ratio $U/t$. We have probed that for SL's with $L_{\rm U}\geq L_0$, this AF ordering is long ranged at $T=0$ but the correspondent order parameter decreases for large interaction values. In fact, the exact dependence of this order parameter with the strength of the interactions depends non-trivially on the superlattice configuration. In turn, some thermodynamical properties, e.g., the temperatures in which spin and charge-fluctuations associated with AF and moment formation start to develop, can be described by a model of an effective homogeneously distributed $U$. The SL's have a dominant short-ranged AF ordering at finite temperatures regardless of their different layer's construction whose onset follows an RPA-like form: $T \propto \exp[-2\pi \sqrt{t/U_{\rm eff}}]$. This is confirmed by the position of the peak in magnetic susceptibility, as well as, in the low-$T$ peak for the specific heat. Regarding the charge dynamics, the kinetic energy clearly shows an anisotropic behavior, where transport preferentially takes place in the direction parallel to the layers. These results suggest a mechanism of reduced dimensionality induced by the increasing interactions in a layered pattern. In the large $U/t$ limit, this would ultimately result in a decoupling of the repulsive and free layers (or strips). Whether this leads to a transition from a Mott-insulator to an anisotropic metal is still an open question 
that deserves further investigation. In connection with the cooling problem in optical lattices, we have also showed a potential cooling protocol where one can more than halve the temperature of the system by adiabatically switching on the interactions in some layers of the lattice. This may renew interest in cooling mechanisms that could eventually reach temperatures to realize the long-sought after $d$-wave superconductivity in cold atoms experiments.

\begin{acknowledgments}
We are indebted to T.~Mendes-Santos, R.~R. dos Santos and R.~T. Scalettar for useful discussions.
RM is financially supported by the National Natural Science Foundation of China (NSFC) (Grant Nos. U1530401, 11674021 and 11650110441). TP gratefully acknowledges financial support from the Brazilian Agencies CNPq and  FAPERJ as well as the  INCT on Quantum Information. We acknowledge the use of computational facilities at CENAPAD-SP and in the Tianhe-2JK at the Beijing Computational Science Research Center (CSRC).
\end{acknowledgments}

\bibliography{sl_bibliography}

\end{document}